\documentclass[preprint,onecolumn,nofootinbib]{revtex4}
\pdfoutput=1
\usepackage[colorlinks=true,linkcolor=blue,urlcolor=blue,filecolor=black,citecolor=red,pdfstartview=FitV,pdftitle={},pdfsubject={},pdfkeywords={},pdfpagemode=None,bookmarksopen=true]{hyperref}
\usepackage{graphicx}
\usepackage{amsmath}
\usepackage{amsfonts}
\usepackage{amssymb,ulem}
\usepackage{color}%
\usepackage{tikz}
\usepackage{dcolumn}
\usepackage{xcolor}

\newcommand{\pd}[3]{\left. \frac{\partial #1}{\partial #2} \right|_{#3}}

\begin{document}
\title{Entanglement of Purification in Holographic Systems}
\author{Peng Liu $^{1}$}
\email{phylp@jnu.edu.cn}
\author{Yi Ling $^{2,3}$}
\email{lingy@ihep.ac.cn}
\author{Chao Niu $^{1}$}
\email{niuchaophy@gmail.com}
\author{Jian-Pin Wu $^{4}$}
\email{jianpinwu@yzu.edu.cn} \affiliation{
$^1$ Department of Physics and Siyuan Laboratory, Jinan University, Guangzhou 510632, China\\
$^2$ Institute of High Energy Physics, Chinese Academy of Sciences, Beijing 100049, China\\
$^3$ School of Physics, University of Chinese Academy of Sciences, Beijing 100049, China\\
$^4$ Center for Gravitation and Cosmology, College of Physical
Science and Technology, Yangzhou University, Yangzhou 225009,
China
}

\begin{abstract}

The holographic entanglement of purification (EoP) in AdS$_{4}$ and AdS-RN black hole backgrounds is studied. We develop an algorithm to compute the EoP for bipartite configuration with infinitely long strips. The temperature behavior of EoP is revealed for small, intermediate and large configurations: EoP monotonically increases with the temperature for small configurations; while for intermediate configurations, EoP is configuration-dependent; EoP vanishes for large configurations. Our numerical results verify some important inequalities of EoP, which we also prove geometrically in Poincar\'e coordinate.

\end{abstract}
\maketitle
\tableofcontents
\section{Introduction}
Quantum entanglement, as a typical phenomenon of quantum system, has been widely studied in quantum information and condensed matter physics since it is powerful in characterizing quantum phase transitions involving strong correlations or topological order \cite{Osterloh:2002na,Amico:2007ag,Wen:2006topo,Kitaev:2006topo}. 
Recent studies also revealed that quantum entanglement plays a key role in understanding spacetime emergence from a holographic viewpoint
\cite{Ryu:2006bv,Hubeny:2007xt,Lewkowycz:2013nqa,Dong:2016hjy}. 
Quantum entanglement has been becoming the core of the interdiscipline of quantum information, condensed matter physics and quantum gravity.

Information-related quantities, such as entanglement entropy (EE), are usually extremely difficult to compute when the degree of freedom is large.
Remarkably, gauge/gravity duality provides an elegant geometric prescription of quantum entanglement.
The entanglement entropy of the sub-region on the boundary was proposed to be proportional to the area of the minimum surface stretching into the bulk of the dual spacetime \cite{Ryu:2006bv}. The holographic entanglement entropy (HEE) can diagnose holographic phase transitions, which is one of the most important applications of HEE \cite{Nishioka:2006gr,Klebanov:2007ws,Pakman:2008ui,Zhang:2016rcm,Zeng:2016fsb}.

Although EE is widely accepted as a good measure to characterize the entanglement of a pure state, it is not suitable for characterizing the entanglement of mixed states. Many new measures have been proposed to characterize mixed state entanglement, such as the non-negativity, entanglement of purification and the entanglement of formation \cite{vidal:2002,Horodecki:2009review}. Mixed states are ubiquitous in both nature and holographic systems. For instance, a thermal quantum system dual to a black hole system is described by a mixed state. It is desirable to study the entanglement properties of black hole systems by means of mixed state entanglement measures.

Recently the entanglement of purification (EoP) was proposed to be proportional to the area of the minimum cross-section of the entanglement wedge \cite{Takayanagi:2017knl,Nguyen:2017yqw}. This prescription provides a novel tool for the study of the mixed state entanglement in holographic theory. Recent progress on the holographic EoP can be briefly reviewed as follows. {To support this proposal, EoP in AdS$_{3}$ and $3$-d BTZ black hole was originally analyzed} in \cite{Takayanagi:2017knl}. 
EoP can be computed analytically in both cases because only symmetrical configurations are considered.
{In the case of AdS${_3}$, the general configuration EoP can be derived by conformal map; while in the case of $(2 + 1)$-d planar BTZ black hole system, the special configuration EoP with $A\cup B = \text{boundary}$ is considered, where the minimum cross-section is relatively straightforward.} 
Since BTZ black hole is the quotient spacetime of AdS$_{3}$, the general configuration EoP in general coordinates can be analytically solved by conformal map. 
The holographic prescription satisfies all relevant inequalities of EoP, which indicates that the minimum cross-section is indeed a good candidate for the holographic EoP.
{The multi-party EoP was subsequently studied in \cite{Umemoto:2018jpc}, where the system was restricted to a symmetrical configuration to simplify the calculation. EoP for a symmetrical configuration was also studied in the quenched system \cite{Yang:2018gfq}.}
More recently, the EoP has also been studied from the viewpoint of dual density matrix, entanglement wedge reconstruction and holographic bit thread \cite{Hirai:2018jwy,Tamaoka:2018ned,Espindola:2018ozt,Agon:2018lwq,Guo:2019azy,Ghodrati:2019hnn}.

The general configuration EoP is not yet fully investigated, and is therefore more desirable to study than symmetrical configuration EoP. The main reason is that it is difficult to locate the minimum cross-section in general configurations.
There are two obstacles in calculating EoP for general configurations.
First, a group of highly non-linear partial differential equations must be solved to locate a minimum surface in a gravitational system, which is often hard to address.
Second, it is often burdened with massive calculation to locate the minimum cross-section in the entanglement wedge.
One way to simplify the calculation is by focusing only on homogeneous backgrounds.
In recent years, homogeneous backgrounds have been studied extensively in the holographic approach.
In addition, one can focus only on general but simple configurations, such as the infinite strips, where the minimum surface can be obtained by solving ordinary differential equations.

In this paper, we study the EoP of bipartite infinite strips in AdS$_{4}$ and AdS-RN black hole background. 
We design an efficient algorithm to numerically calculate the EoP for general configurations, by using the symmetry and nature of the system and the EoP.
First, the bipartite EoP in AdS$_{4}$ spacetime is fully studied {by taking advantage of the global scaling symmetry, which means that we fully reveal the small configuration EoP properties of any background with asymptotic AdS$ {_4}$.}
Second, the EoP behaviors with temperature for small, intermediate and large configurations are discussed for AdS-RN black hole: EoP monotonically increases with temperature for small configurations; the temperature behavior of EoP depends on the details of configurations for intermediate configurations; EoP vanishes as MI vanishes for large configurations and high temperature limit. Numerical results in this paper also verify some important inequalities of EoP, which we also prove in Poincar\'e coordinate in geometric manner.

The paper is organized as follows: we introduce the concept of entanglement of purification and its holographic duality {in section \ref{alg}}. In particular, we develop an algorithm to calculate the EoP for bipartite infinite strip configuration in homogeneous backgrounds. We then study the EoP behaviors for pure AdS$_{4}$ spacetime in section \ref{ads}, AdS-RN black hole systems in \ref{adsrn}. {Our conclusion and discussion is given in section \ref{discussion}, and we provide geometrical proofs of some inequalities of EoP in appendix \ref{sec:geoprof}.}

\section{The minimum surface for infinite strip partition}\label{alg}
First we introduce the concept and the holographic duality of EoP. We then develop an algorithm to calculate holographic EoP for bipartite strips by using the EoP's geometric interpretation.

\subsection{Holographic Entanglement of Purification}
One of the most striking features of quantum mechanics is that 
subsystems can entangle with each other.
Especially, for a pure state $|\psi\rangle$ composed of two sub-systems $A$ and $B$, the entanglement between $A$ and $B$ can be captured by observers who have only access to $A$ or $B$. 
{The subsystem $ A $ behaves as a reduced matrix $\rho_{A} = \text{Tr}_{B} \left(|\psi\rangle\langle\psi|\right)$ for observers  constrained to $A$}. The mixed property of $\rho_{A}$ comes from the entanglement between $A$ and $B$. A natural quantity to measure this entanglement is the von Newmann entropy of $\rho_{A}$,
\begin{equation}\label{ee-von}
    S_{A} (|\psi\rangle) = - \text{Tr}\left[ \rho_{A} \log \rho_{A} \right],
\end{equation}
which is dubbed as the entanglement entropy (EE). Note that for pure states $S_{A} = S_{B}$ \cite{Chuang:2002book}. 
The entanglement entropy $ S $ in field theory diverges with the area law due to the divergences from the UV degree of freedom.
Regularization is therefore necessary to achieve a final EE for field theory.
Given the definition of HEE, it is then readily to define the mutual information (MI),
\begin{equation}\label{mi:def}
 I\left(A,B\right) := S\left(A\right) + S\left(B\right) - S\left(A\cup B\right),
\end{equation}
which measures the entanglement between two separate subsystems $A$ and $B$. It is clear that $ \rho_{AB} = \rho_{A} \otimes \rho_{B} $ when $ I\left(A,B\right) =0 $. Moreover the MI is always finite since the divergence in EE is canceled out.

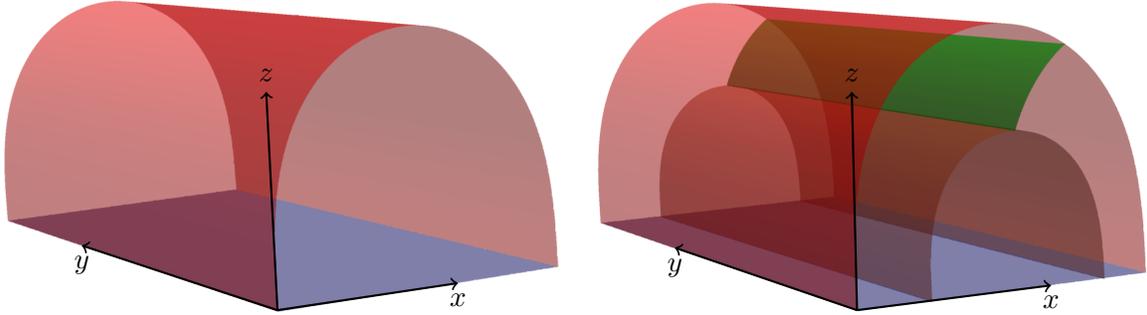
\begin{figure}
\begin{tikzpicture}[scale=1]
\node [above right] at (0,0) {\includegraphics[width=7.5cm]{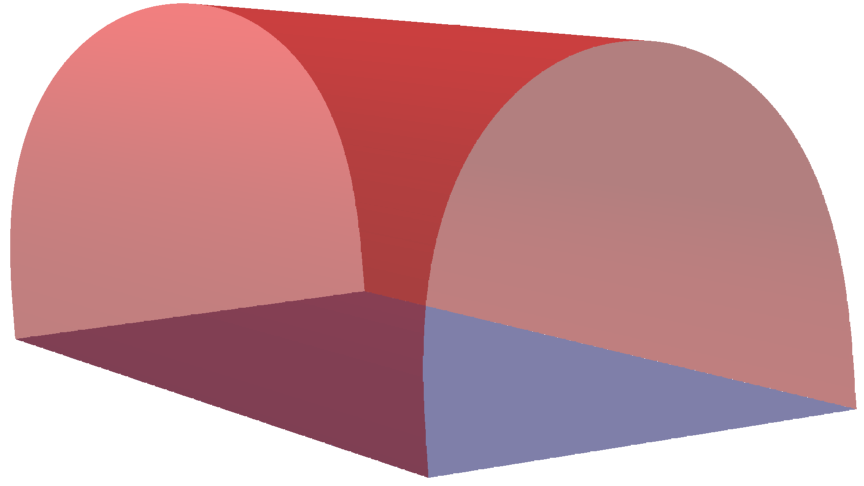}};
\draw [right,->,thick] (3.85, 0.22) -- (6.25, 0.58) node[below] {$x$};
\draw [right,->,thick] (3.85, 0.22) -- (1.25, 1.08) node[below] {$y$};
\draw [right,->,thick] (3.85, 0.22) -- (3.7, 3.125) node[above] {$z$};
\end{tikzpicture}
\begin{tikzpicture}[scale=1]
\node [above right] at (0,0) {\includegraphics[width=7.5cm]{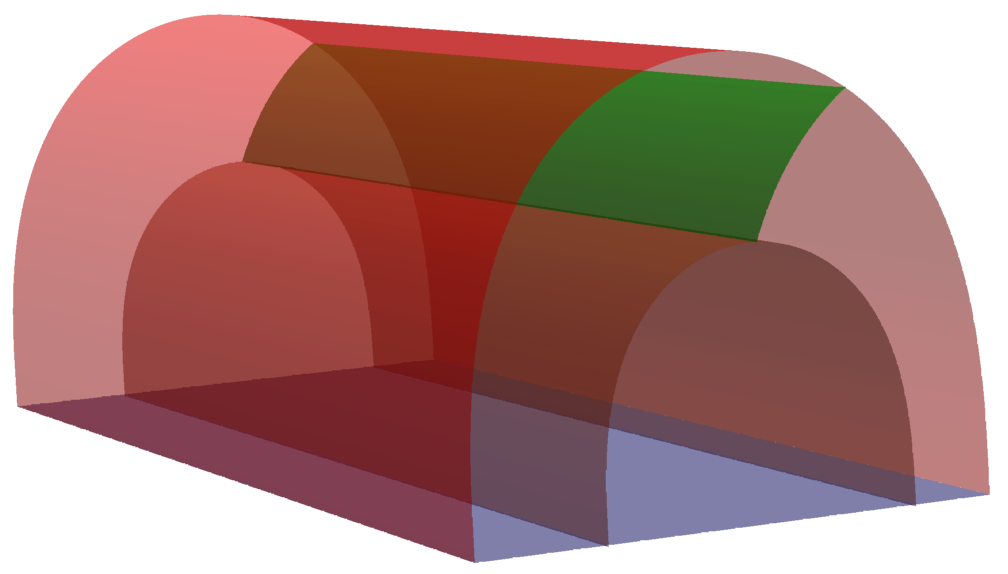}};
\draw [right,->,thick] (3.67, 0.22) -- (6.25, 0.55) node[below] {$x$};
\draw [right,->,thick] (3.67, 0.22) -- (1.25, 1.05) node[below] {$y$};
\draw [right,->,thick] (3.67, 0.22) -- (3.6, 3.125) node[above] {$z$};
\end{tikzpicture}

\caption{The left plot: The minimum surface for a given width $w$. The right plot: The minimum cross-section (green surface) of the entanglement wedge.}
\label{msd1}
\end{figure}

Entanglement entropy can describe pure state entanglement, but is not suitable for characterizing the mixed state entanglement. The reason is that, not only the entanglement property but also the mixed property contributes to the entanglement entropy for mixed states.
For example, the entanglement entropy for a product state $\rho_{A}\otimes \rho_{B}$, where the degrees of freedom in $ A $ and $ B $ do not entangle, can be non-zero. 
{Many new measures to diagnose the mixed state entanglement have been proposed \cite{vidal:2002,Horodecki:2009review}.} The EoP is one of the useful measures for mixed state entanglement, which involves the purification of mixed states. A mixed state $\rho$ on $\mathcal H_{A} \otimes \mathcal H_{B}$ can be purified by introducing extra degrees of freedom $A'$ (entangled with $A$) and $B'$ (entangled with $B$) such that $\rho$ arises as the reduced matrix from a pure state $|\psi\rangle \in \mathcal H_{AA'} \otimes \mathcal H_{BB'}$. Obviously there exists infinite ways to purify $\rho$, and the EoP $E_{p}\left(\rho\right)$ is defined as \cite{Terhal:2002}
\begin{equation}\label{eop:def}
    E_{p}(\rho) := \min_{|\psi\rangle: \rho = \text{Tr}_{A'B'} |\psi\rangle \langle \psi|} S_{AA'}(|\psi\rangle).
\end{equation}
{The Eq. \eqref{eop:def} shows that the entanglement of purification involves a double minimization procedure, over all possible purifications and all possible bipartitions of the extra degrees of freedom.}
EoP can measure both quantum correlation and classical correlation of two sub-regions \cite{Terhal:2002}. 
{EoP satisfies several important inequalities. Therefore, its correct holographic dual must also satisfy these inequalities \cite{Takayanagi:2017knl,Bao:2017nhh}.}

{The HEE (see the left plot of Fig. \ref{msd1}) was proposed as the area of the minimum surface in dual gravity systems \cite{Ryu:2006bv}.}
The success of HEE has prompted experts to study the geometrical duality of more information-related physical quantities, which greatly simplifies the study of the  quantum information in strongly correlated systems.
Takayanagi proposed that the EoP $ E_{W}\left(\rho_{AB}\right) $ is associated with a minimum cross-section $ \Sigma_{AB} $ in connected entanglement wedge \cite{Takayanagi:2017knl}, {\it i.e.}, the configurations with non-zero MI (see the right plot in Fig. \ref{msd1}),
\begin{equation}\label{heop:def}
    E_{W}\left(\rho_{AB}\right) = \min_{\Sigma_{AB}} \left( \frac{\text{Area} \left(\Sigma_{AB}\right)}{4G_{N}}\right) .
\end{equation}
EoP vanishes for configurations with disconnected entanglement wedge (zero MI). The prescription of the EoP with minimum cross-section indeed satisfies all existing inequalities of EoP \cite{Takayanagi:2017knl}.

The EoP computation depends on the MI and entanglement wedge, both related to the minimum surface. Therefore, we discuss how to locate the minimum surface for infinitely long strip on the boundary by Euler-Lagrange method.

\subsection{Computations of minimum surface with arc length parameter}
We start with a generic homogeneous background
\begin{equation}\label{genbg}
ds^{2} = {g_{tt}} dt^2 + g_{zz}dz^2 + g_{xx}dx^2 + g_{yy} dy^2,
\end{equation}
with $z=0$ representing the asymptotic AdS boundary \footnote{The numerical method we will show next is also applicable to metric with off-diagonal components.}. The homogeneity means that all metric components $g_{\mu\nu}$ are only functions of $z$.

The left plot in Fig. 1 shows a cartoon of the minimum surface for an infinitely strip.
The area of the minimum surface is given by
\begin{equation}\label{mse}
A_\Sigma = \iint \sqrt{g_{yy} \left( g_{xx} dx^{2} + g_{zz} dz^{2} \right)} dy= \iint\sqrt{g_{yy}\left(g_{xx} + g_{zz} z'(x)^2\right)}dxdy.
\end{equation}
{Note that the minimum surface is invariant along the direction of $y$, so one can integrate out $y$ and calculate the minimum surface for a one-dimensional system.}
As a result, we can transform \eqref{mse} into
\begin{equation}\label{reduce-y}
    A_\Sigma = L_y \int_0^w \sqrt{g_{yy}\left(g_{xx} + g_{zz} z'(x)^2\right)} dx,
\end{equation}
where $ \tilde L_{y} = \int dy $, and the width of the strip $ \tilde w=\int dx $. Hence the minimum surface can be described by $z\left(x\right)$. 
{From now on, we will denote the minimum surface as $z(x)$ and call it the minimum curve or geodesic.}
Ignoring several common factors, we label the EE as $ \tilde S_{A} \equiv \int_0^w \sqrt{g_{yy}\left(g_{xx} + g_{zz} z'(x)^2\right)} dx$ for convenience.
It is worth noting that the asymptotic AdS boundary will result in a common divergence in the HEE. We subtract this divergence to retrieve finite results of HEE. Treating \eqref{reduce-y} as an action, the geodesic is given by the solution of the Euler-Lagrange equation,
\begin{equation}\label{eom1}
    2 g_{{yy}} g_{{zz}} z'(x)^2 g_{{xx}}'+g_{{xx}} \left(g_{{yy}} \left(-2 g_{{zz}} z''(x)-z'(x)^2 g_{{zz}}'+g_{{xx}}'\right)+g_{{zz}} z'(x)^2 g_{{yy}}'\right)+g_{{xx}}^2 g_{{yy}}'=0,
\end{equation}
where $g_{\#\#}' \equiv g'_{\#\#}(z)$. Eq. \eqref{eom1} usually requires numerical treatments due to its high non-linearity. Given $z(0)=\epsilon$ and $z'(0)$, a numerical solution can be obtained by {\texttt{NDSolve}} with Mathematica. With the solution $z(x)$, it is readily to obtain the width $\tilde w$ of the strip. In addition, the arc length parameter $s(x)$ can be obtained by integrating the $A_{\Sigma}$ from $x=0$ to $x$.

The above method involves time-consuming numerical integration. Alternatively, \eqref{eom1} can be solved by treating it as a two-variable system with arc length parameter $s$,
\begin{eqnarray}
&g_{{xx}} g'_{{yy}}+g_{{yy}} \left(g'_{{xx}}-2 g_{{xx}} g_{{zz}} z'^2 g'_{{yy}}\right)-g_{{yy}}^2 \left[g_{{xx}} z'^2 g'_{{zz}}+g_{{zz}} \left(z'^2 g'_{{xx}}+2 g_{{xx}} z''\right)\right]=0, \label{eom_arc:1} \\
&g_{xx} x'(s)^2 + g_{zz} z'(s)^2 - g^{-1}_{yy}=0. \label{eom_arc:2}
\end{eqnarray}
Again, the $g'_{\#\#}\equiv g'_{\#\#}(z)$, but $z''\equiv z''(s),z'\equiv z'(s),x'\equiv x'(s)$ represent derivatives with respect to arc length parameter $s$.
EOMs \eqref{eom_arc:1}-\eqref{eom_arc:2} can be derived from
\begin{eqnarray}
&A_\Sigma = L_y \int_0^w \sqrt{g_{yy}\left(g_{xx} x'(s)^2 + g_{zz} z'(s)^2\right)} ds, \label{lag_arc:1}\\
&g_{yy}\left(g_{xx} x'(s)^2 + g_{zz} z'(s)^2\right) =1, \label{lag_arc:2}
\end{eqnarray}
where \eqref{lag_arc:2} is the constraint from setting $s$ as the arc length parameter. In this way, the time-consuming numerical integration is unnecessary.

Once the geodesic is solved, we are ready to calculate the EoP with the area of minimum cross-section.

\subsection{Computations of holographic EoP}\label{eopcomp}
\begin{figure}[htbp]
    \centering
    \includegraphics[width=0.5\textwidth]{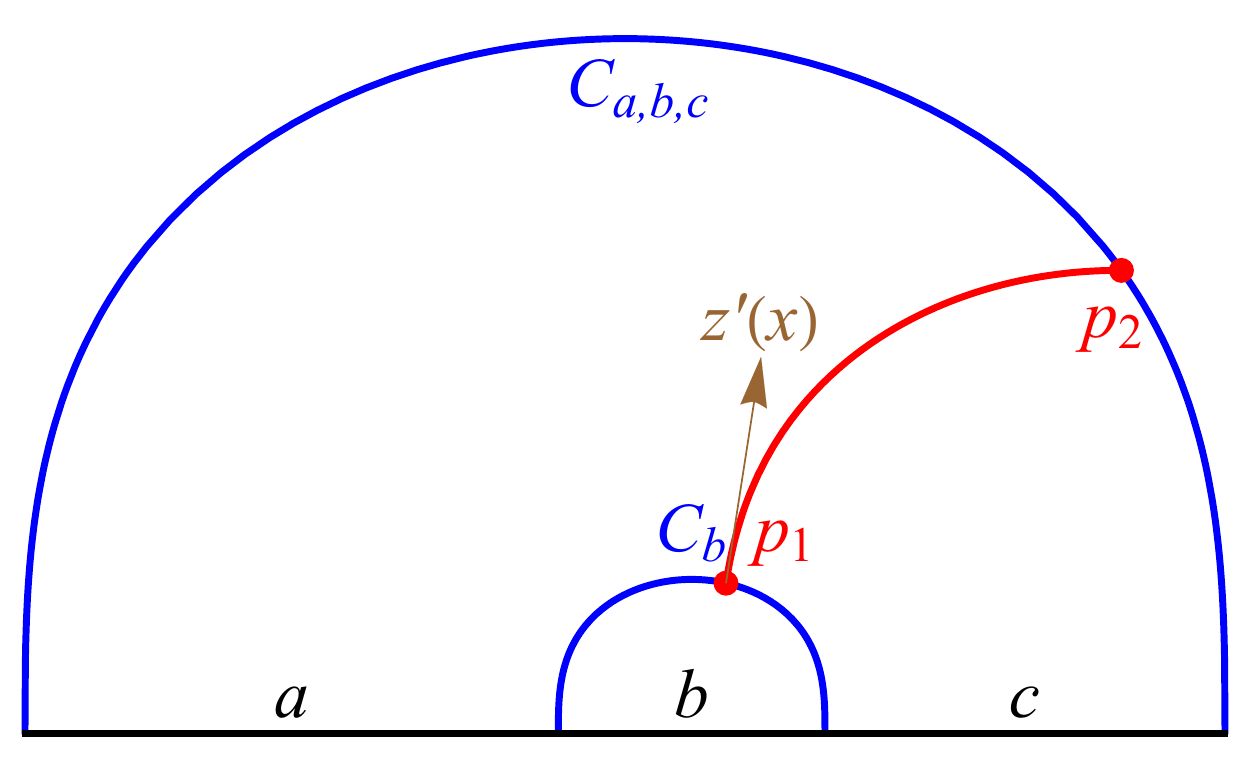}
\caption{One bipartite configuration can be specified by three parameters $(a,b,c)$, where $a$ and $c$ are the width of two infinitely long strips,respectively, while $b$ is the separation. $C_{\#\#}$ represents the minimum surface ending on configuration $\# \cup\#$. The red curve $z(x)$ is the geodesic connecting $p_{1}$ and $p_{2}$, and the brown arrow represents the tangent vector $z'(x)|_{p_{1}}$.}
    \label{fig:config}
\end{figure}

Given a configuration $\left(a,b,c\right)$ with non-zero MI (see Fig. \ref{fig:config}), the EoP corresponds to the length of the minimum geodesic. {In order to locate the minimum geodesic, the first step is to find the geodesic connecting connecting $ p_{1}\in C_{b}$ and $ p_{2} \in C_{a,b,c} $. This can be obtained by the following method.} Given a slope $z'(x)|_{p_{1}}$ {at $ p_{1} \in C_{b} $}, a unique geodesic can be obtained by solving \eqref{eom_arc:1}-\eqref{eom_arc:2}. For a large enough $z'(x)|_{p_{1}}$, the geodesic intersects with $C_{a,b,c}$ at $p_{2}$, and the length $l(p_{1},p_{2})$ between $p_{1}$ and $p_{2}$ can be read off as $|s(p_{2})-s(p_{1})|$.\footnote{The arc length parameter $ s(p) $ of $ p \in (z(s),x(s)) $ can be obtained by solving $ s $ from $ x(s) = x|_p $.} 
{The EoP can thus be obtained by searching for the minimum value of $l(p_{1},p_{2})$ in space $(p_{1}\in C_{b},p_{2}\in C_{a,b,c})$, or equivalently in space $ (p_{1}\in C_{b},z'(x)|_{p_{1}}) $. From the above method we can see that EoP calculation is hard, because it needs to search for the minimum value in 2-d space, and each search requires cumbersome calculations.}

\begin{figure}[htbp]
    \centering
    \includegraphics[width=0.45\textwidth]{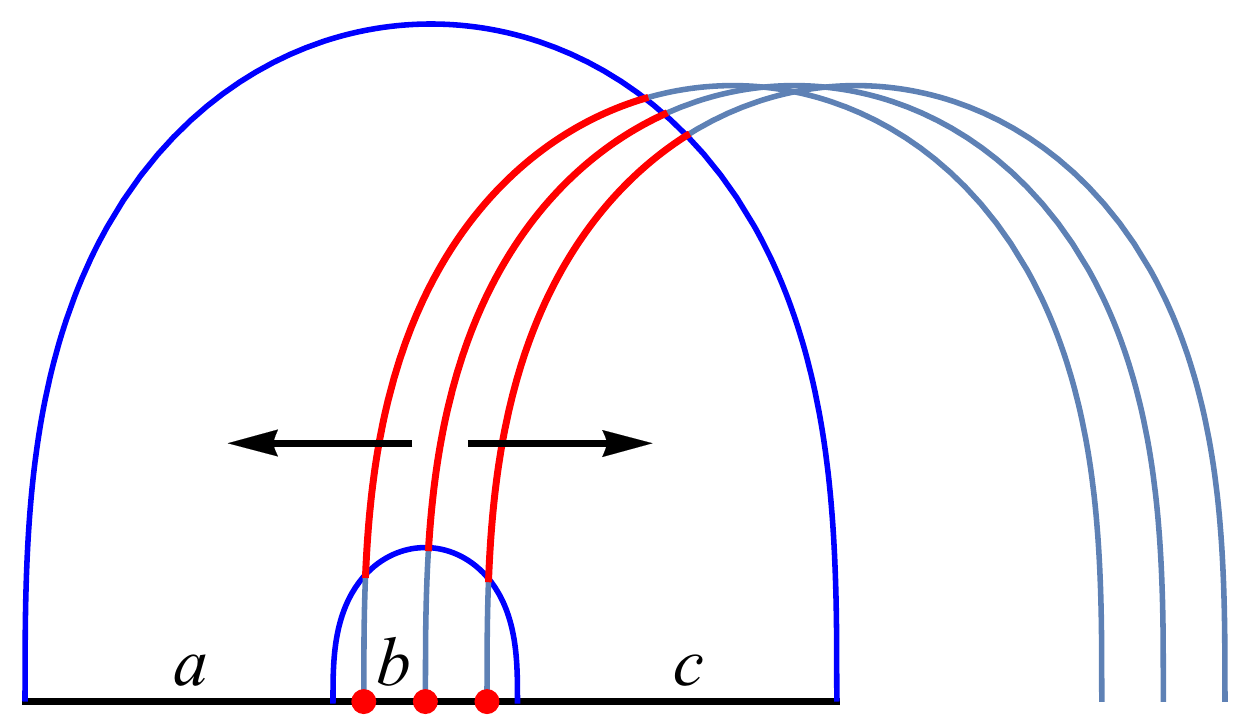}\quad
    \includegraphics[width=0.45\textwidth]{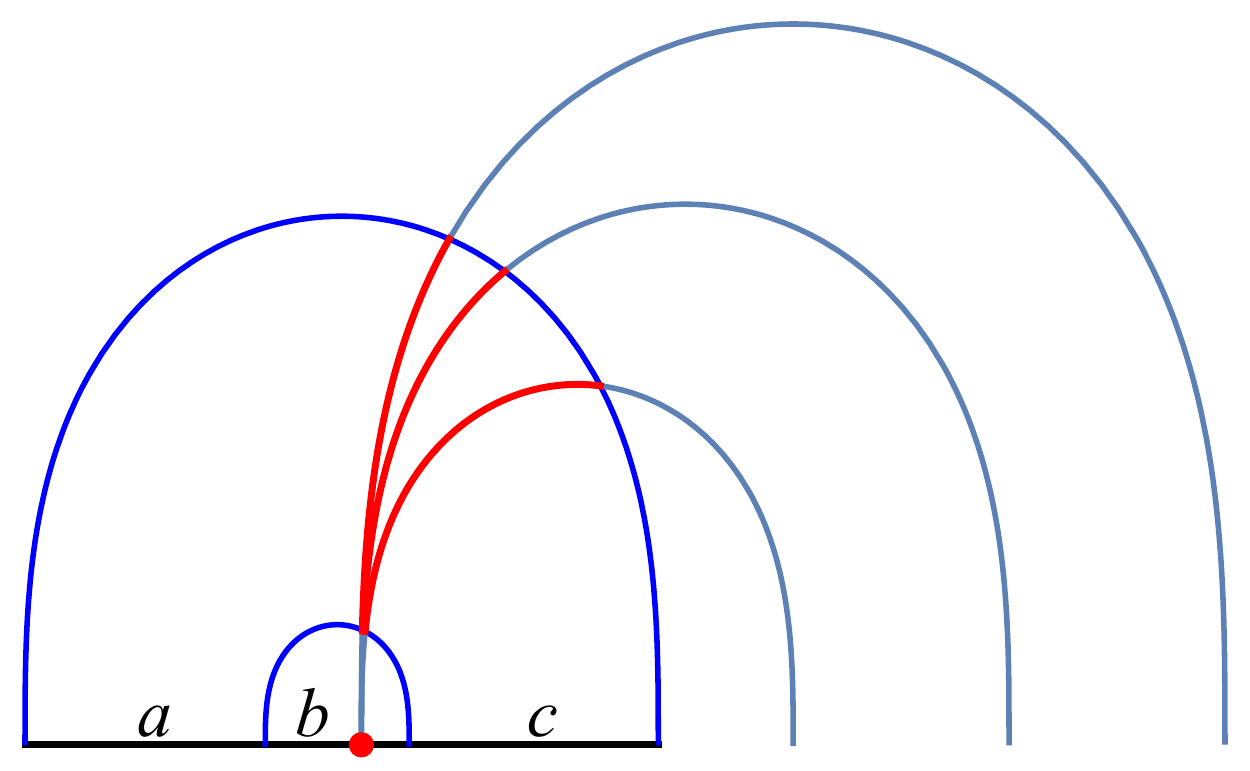}
    \caption{Left plot: a demonstration of shifting a curve to cut the $C_{b}$ and $C_{a,b,c}$. The red dots are endpoints. Right plot: a demonstration of solving curves at different $v=z'(0)$, and use them to cut $C_{b}$ and $C_{a,b,c}$. The red curves in each plot are the minimal segments connecting each pair of intersections.
    }
    \label{fig:algo}
\end{figure}

We present some tricks to speed up the computation of EoP. First, we only need to focus on the area near the bottom of $ C_b $. Due to the singularity of the asymptotic AdS boundary, the region closer to the boundary contributes more to the minimum surface area. Therefore, the minimum cross-section will end only on the region near the bottom of the $ C_{b} $.
This observation is also verified by subsequent numerical computations (see section \ref{ads} and section \ref{adsrn}). Second, the homogeneity of the background and the infinite length of the strip can be used to further narrow the search space. 
{The symmetry shows that the EoP of $(a, b, c)$ is equal to that of $(c, b, a)$, so that we only need to calculate the situation for $a > c$.}
Moreover, a necessary condition for non-zero MI is $\left(a>b\right)\wedge
\left(c>b\right)$ \footnote{If $ a<b | c < b $ then the MI will be zero since $ l_{C_{a,b,c}} > l_{C_{a}}, \, l_{C_{a,b,c}} > l_{C_{b}}$. {Note also that this holds only for Poincar\'e patch.}}. Given the above considerations, we narrow down the search space to,
\begin{equation}\label{config1}
    \left(a>b\right)\wedge \left(c>b\right) \wedge \left(a>c\right).
\end{equation}
{Furthermore, the homogeneity of the background guarantees that a geodesic is still geodesic after a translation. This fact can be used to significantly reduce the amount of computation, because we can get another geodesic by translating a geodesic, without having to re-solve the equations of motion (see the left plot of Fig. \ref{fig:algo}).}
Furthermore, for a background like AdS$_{d}$ spacetime, scaling symmetries can be used to greatly simplify the numerical computation of EoP, which we will elaborate in section \ref{ads}.

\begin{figure}[htbp]
	\centering
	\includegraphics[width=0.6\textwidth]{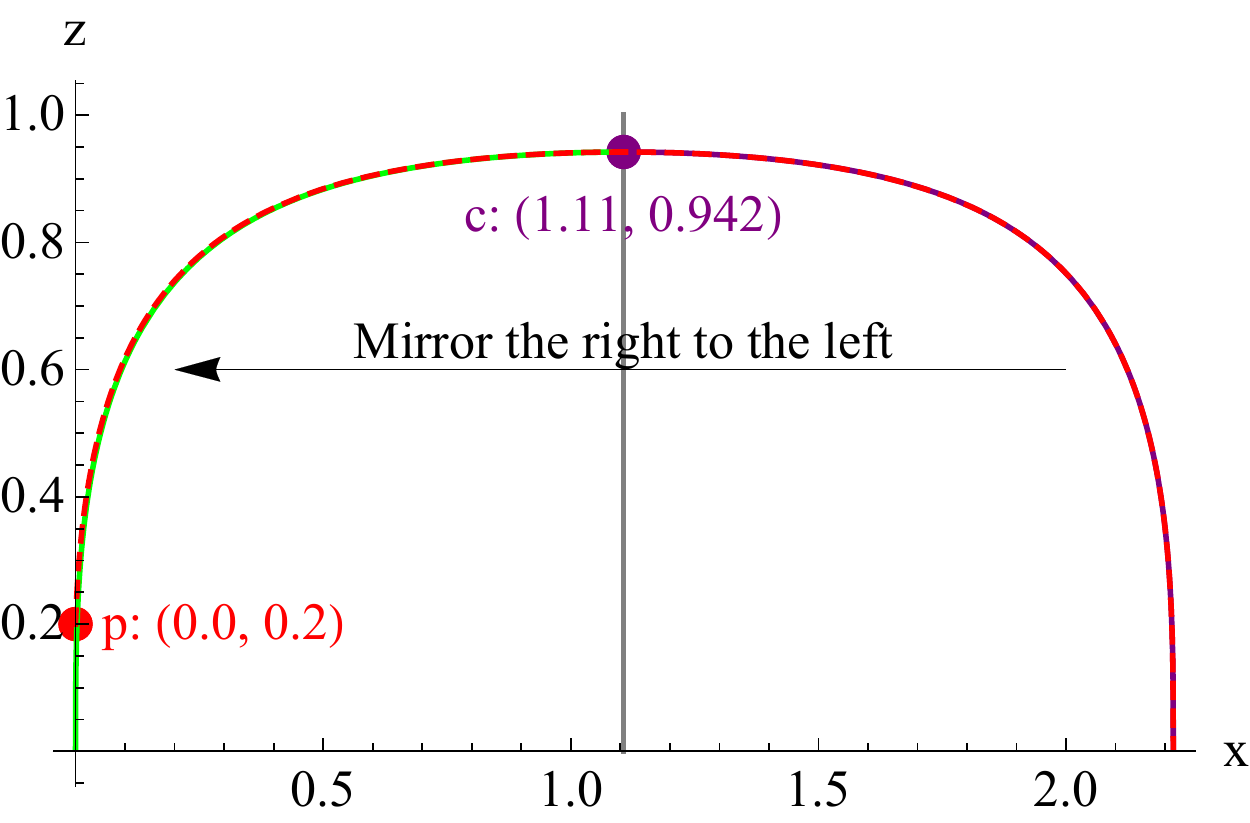}
	\caption{{The cartoon of obtaining the full $ z(x) $. The red dashed curve is the minimal curve started from $ p = (0,0.2) $ (red dot) with $ z'(x) = 10637.3 $. The bottom point is at $ c=(1.11,0.942) $ (purple dot), and the purple dashed line is the segment from $ c $ to the right endpoint. The green curve is the mirror image with respect to the gray vertical line through $ c $. Therefore, the union of the green line and the purple line is the full solution of $ z(x) $. }}
	\label{fig:zxsol}
\end{figure}

Since the area near the AdS boundary is divergent, we take a cut off at $z=\epsilon$ at which the geodesic ends. A geodesic $C(x,z')$ can then be specified by $(x(s),z'(s))$ with $z(s)=\epsilon$. 
{Note that the cutoff can not be too small because it can lead to an large $z'(x)$, making numerical findings unreliable. A better approach is to set a finite $ \epsilon $, but we can only get a fragment of the minimum surface in this way. To get the full solution of $z(x)$, we can solve eom in a large range of $s$ to get $\left(x(s),z(s)\right)$. Setting a large range of $s$ ensures that the other endpoint of the smallest surface is sufficiently close to the AdS boundary. Then, we can find the position of the turning point by solving $z'(s_*)=0$. The minimum surface is symmetrical about the line $x(s_*)$, which allows us to mirror the curve from $s_*$ to the right endpoint and stitch it to the complete solution of $z(x)$. 
Notice that the right endpoint can be arbitrarily close to the boundary, hence we can go beyond the original cutoff $ \epsilon $ and obtain the minimal surface with two endpoints arbitrarily close to the boundary by mirror method. We show the method to solve the complete solution of $z(x)$ as Fig. \ref{fig:zxsol}.}
Finally, we divide the algorithm into the following steps,
\begin{enumerate}
\item Given a background and a configuration $\left(a,b,c\right)$ with non-zero MI, one finds the geodesic $C_{b}$ and $C_{a,b,c}$ with finite cutoff at certain values of $(x(s),z')$.
\item Solve the geodesic $C(x,v)$ with a width $w_{1}>c$ on the boundary such that one endpoint of the curve falls into the region $b$, and then find its intersections with $C_{b}$ and $C_{a,b,c}$ at $p_{1}$ and $p_{2}$. The area of the cross-section is $E(x,v) = |s(p_{2}) - s(p_{1})|$. 
\item Translate $C(x,v)$ along $x$ direction with fixed $v$, and find the local minimum $\mathcal E(v)\equiv\min_{x} E(x,v)$ (see the left plot in Fig. \ref{fig:algo}).
\item Vary $v$ and repeat the last two steps at each $v$ such that the global minimum of $E(x,v)$ can be obtained as,
    \begin{equation}\label{eop:alg}
        E_{W}\left(a,b,c\right) = \min_v \mathcal E(v).
    \end{equation}
See the right plot in Fig. \ref{fig:algo}.
\end{enumerate}

In the subsequent two sections, we apply the above algorithm to explore the property of EoP over AdS$_{4}$ spacetime and AdS-RN black hole.

\section{EoP for AdS$_{4}$}\label{ads}

The EoP of a small configuration is dictated by the asymptotic boundary, therefore it is worthwhile to study the EoP in pure AdS$_4$. 
{Compared with AdS$_3$ where the EoP is available in terms of analytical expression of minimum surface, it is difficult to compute the EoP for AdS${_4}$ analytically because the expressions of analytical geodesics are too complicated for practical use \cite{Tonni:2010pv,Kundu:2016dyk}.}

\begin{figure}
\includegraphics[width=0.55\textwidth]{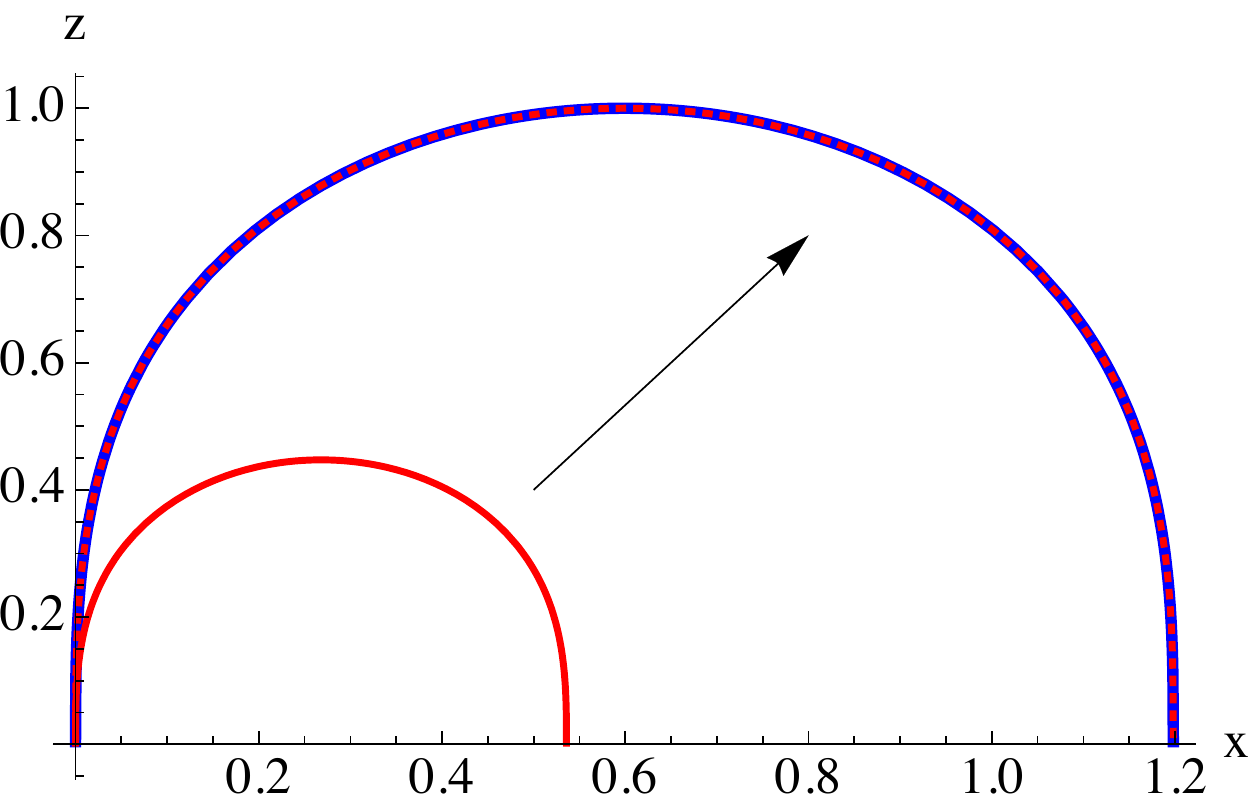}
\caption{The small red curve is $z_{1}(x)$ with $w_{1} = 0.536$, and the large blue curve is $z_{2}(x)$ with $w_{2} = 1.20$. The red dotted curve is $\frac{w_{2}}{w_{1}}z_{1}\left(\frac{w_{1}x}{w_{2}}\right)$, which matches perfectly with $z_{2}(x)$.}
\label{scale}
\end{figure}

Following the algorithm outlined in the previous section, we numerically compute the EoP of AdS$_{4}$. In pure AdS$_{4}$ the equation of motion for the minimum surface reads
\begin{equation}\label{adseom}
    z(x) z''(x)+2 z'(x)^2+2=0.
\end{equation}
Note that the above equation is invariant under $x\to \lambda x, z\to \lambda z$ due to the global scaling symmetry of pure AdS$_{4}$. Therefore $z_{1}(x)$ with width $w_{1}$ can be rescaled to $z_{2}(x)$ with width $w_{2}$ by $z_{2}(x) = w_{2} / w_{1} z_{1}\left(w_{1} x / w_{2}\right)$. This is verified with numerics in Fig. \ref{scale}.
The scaling symmetry of the area of the minimum surface also significantly simplifies the calculation of EoP. The HEE reads as
\begin{equation}\label{adsarea}
S_{\text{AdS}_{4}} = \int_{x_{1}}^{x_{2}} {\frac{\sqrt{1+z'\left(x\right)^{2}}}{z^{2}}dx}.
\end{equation}
It is then readily seen that $S_{\text{AdS}_{4}} \to S_{\text{AdS}_{4}}/\lambda$, and hence $S_{\text{AdS}_{4}}$ has scaling dimension $[ -1 ]$. 
{Thanks to the scaling symmetry of $S_{\text{AdS}_{4}}$, one only needs to solve one curve numerically and then rescale it to any other case.}
Obviously, the scaling symmetry can also simplify the computation of EoP for general AdS$_{d}$.

\begin{figure}
\includegraphics[width=0.6\textwidth]{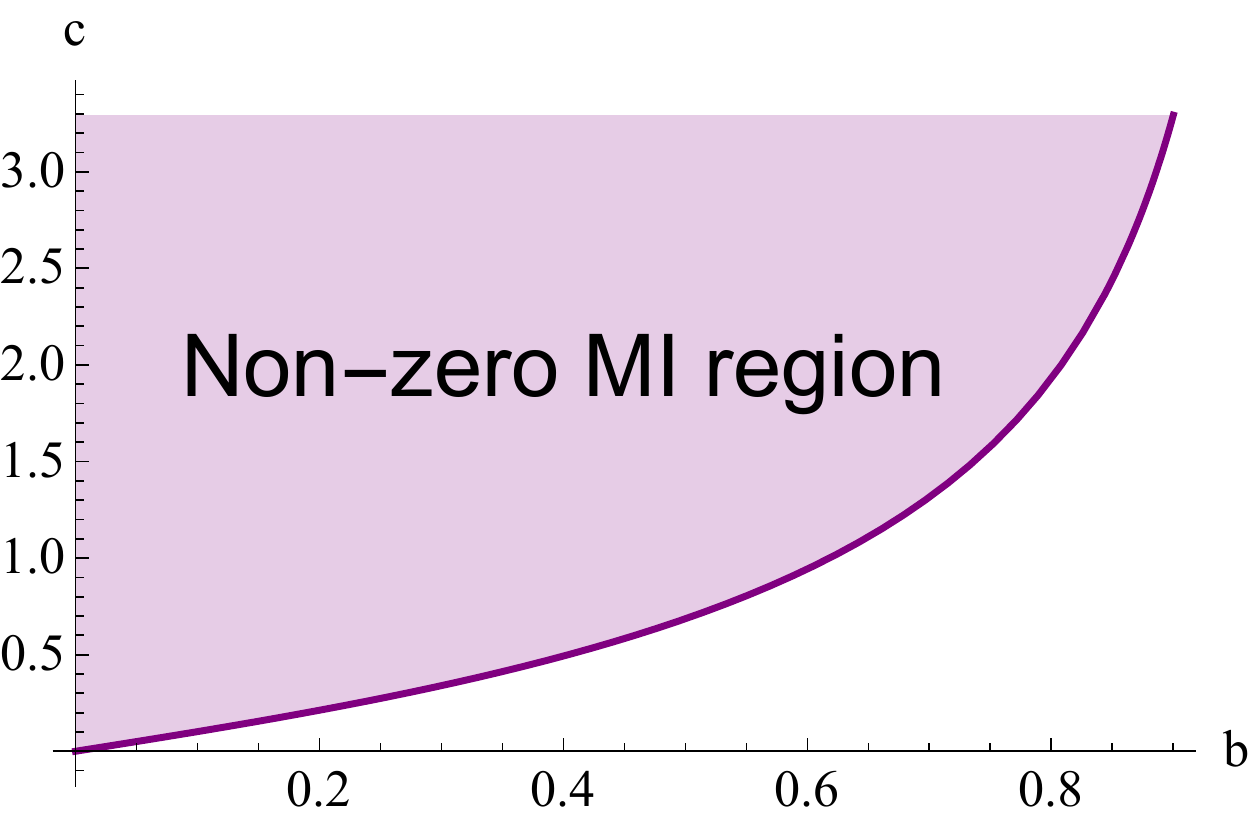}
\caption{
The purple shaded region is the configuration space with non-zero MI for AdS$_{4}$.
}
\label{ntmiregion}
\end{figure}

Subsequently, we deduce the condition for non-zero MI since the EoP is non-zero only when MI is non-zero.
The bipartite configuration can be specified by $(1,b,c)$ due to the scaling symmetry. Using the scaling relation $S_{\text{AdS}_{4}}\sim 1/w$ we see that the non-zero MI requires \footnote{Notice that after subtracting a common divergence $1/\epsilon$ with $\epsilon$ the cutoff, the non-zero part of the HEE is always negative. Hence \eqref{ntmi} is required to obtain the non-zero MI.}
\begin{equation}\label{ntmi}
\frac{1}{c}+1<\frac{1}{c+b+1}+\frac{1}{b}.
\end{equation}
Solving \eqref{ntmi} we obtain
\begin{equation}\label{ntmir}
\left(0<b<1\right)\land \left(c>\frac{1}{2} \sqrt{\frac{b^3-3
b^2-5 b-1}{b-1}}-\frac{1}{2} (b+1)\right).
\end{equation}
Therefore for pure AdS$_{4}$ the EoP is only non-zero in parameter space $(b,c)$ satisfying \eqref{ntmir} (see Fig. \ref{ntmiregion}). It is worth to mention that for more complex systems, such as black hole systems, we may need to directly use numerical calculations to determine the condition for MI non-zero. Next, we explore the details of EoP.

\begin{figure}[htbp]
	\centering
	\includegraphics[width=0.45\textwidth]{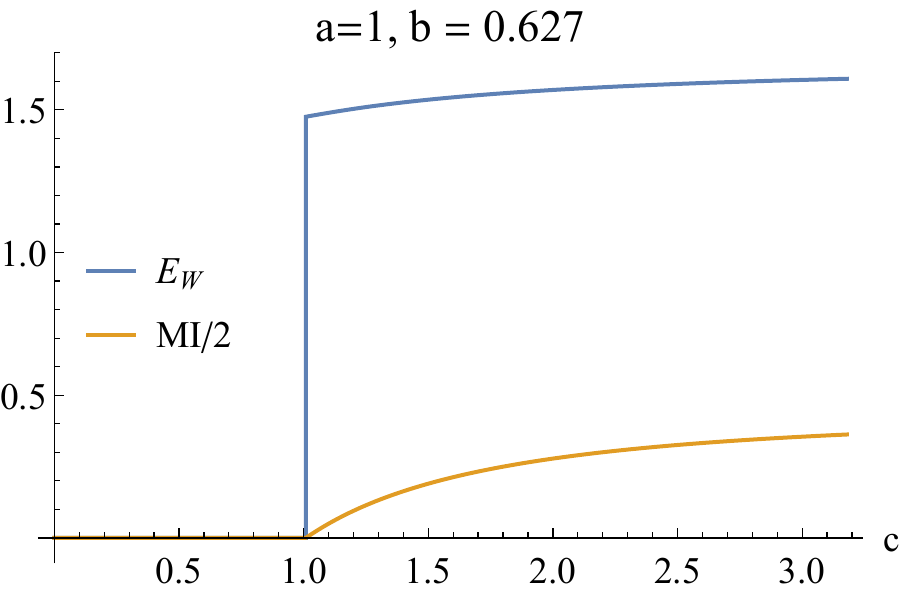}\quad
	\includegraphics[width=0.45\textwidth]{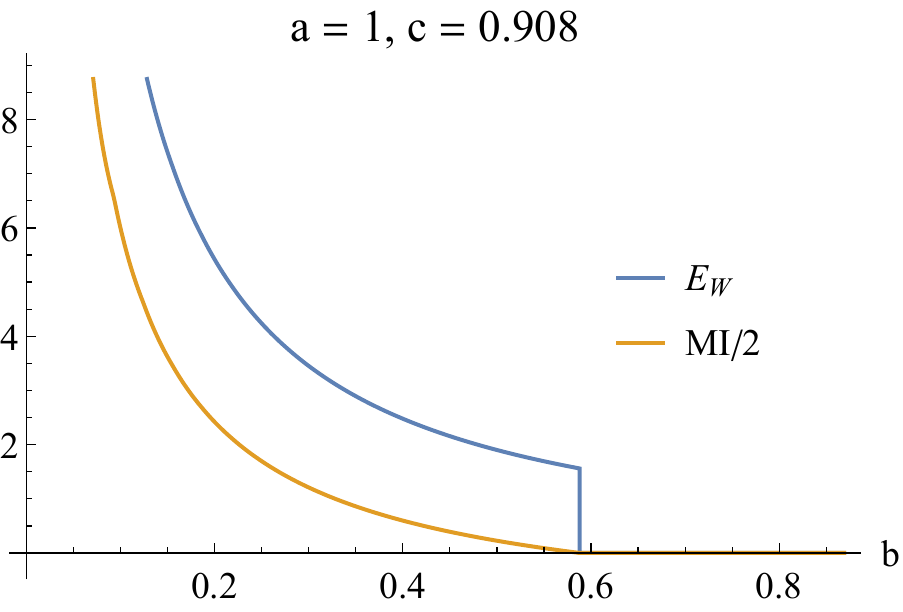}
	\caption{
	Left plot: EoP and MI$ /2 $ along $ c $ at $ (a,b) = (1, 0.627) $. Right plot: EoP and MI$ /2 $ along $ b $ at $ (a,c) = (1, 0.908) $.}
	\label{fig:alongbc}
\end{figure}

We demonstrate the EoP behavior with configurations in Fig. \ref{fig:alongbc}, from which we can see that the EoP increases with $ c $ and decreases with $ b $. This behavior can be understood since the entanglement usually decays with the increase of the separation, and increases with the increase of the size of sub-region. 

We also notice that the MI is continuous, while the EoP undergoes a disentangling phase transition at the point where MI starts to vanish. 
{And we can also see that EoP is always greater than one half of MI. This is actually an important inequality that EoP satisfies \cite{Takayanagi:2017knl}.}

\begin{figure}[htbp]
    \centering
    \includegraphics[width=0.6\textwidth]{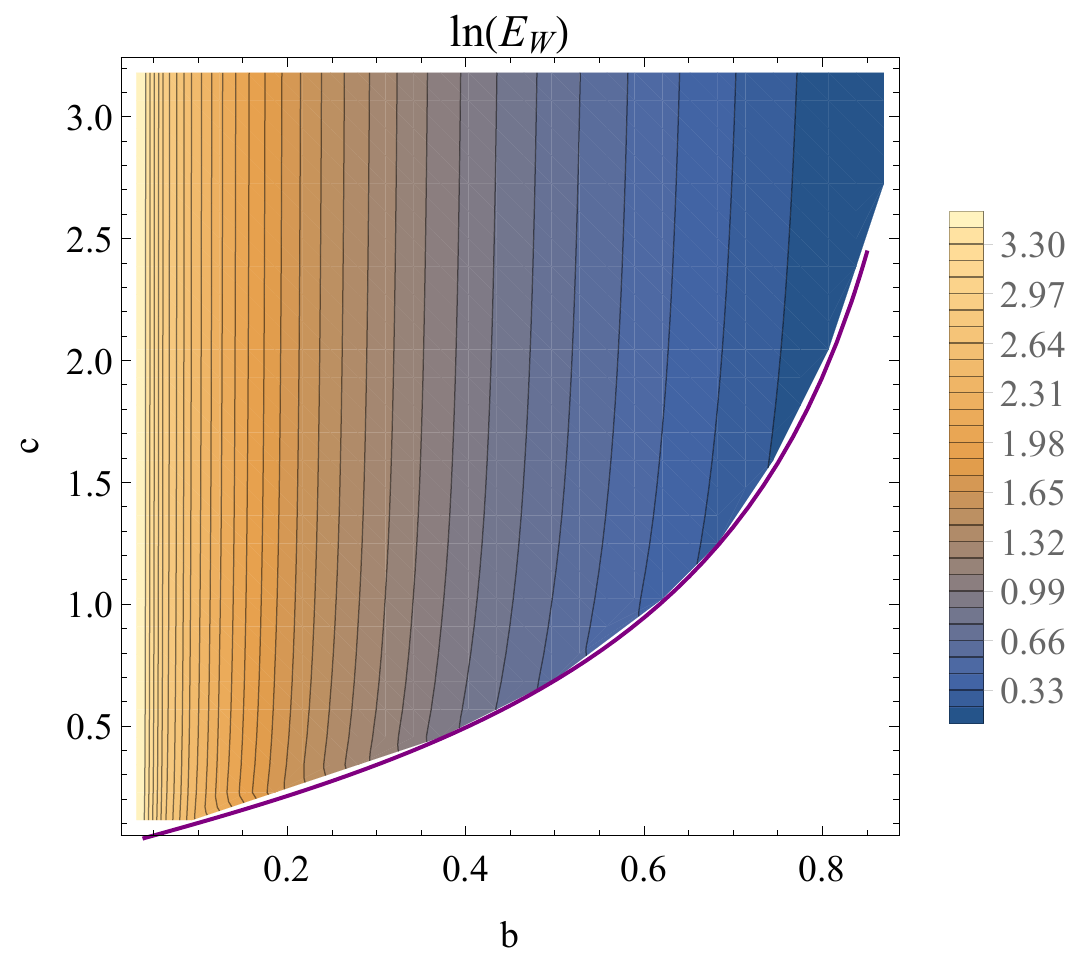}
    \caption{
The contour plot of $\ln E_{W}(b,c)$ over $(b,c)$ space. The purple curve corresponds to the critical line of non-zero MI from the right plot of Fig. \ref{ntmiregion}.
    }\label{figcont}
\end{figure}

{The above phenomena actually reflect three important inequalities of EoP, which we prove in geometrical manner in appendix \ref{sec:geoprof}.}
We also demonstrate the EoP $E_{W}$ over the full parameter space $\left(b,c\right)$ in Fig. \ref{figcont}. 

\section{EoP for AdS-RN Black hole}\label{adsrn}
In this section, we explore the EoP over the background of AdS-RN black hole. First we discuss the EoP computation in AdS-RN black hole. And then we explore the EoP at small, intermediate, and large configurations respectively.

The AdS-RN black hole geometry reads as \cite{Donos:2013eha}
\begin{equation}\label{RNmetric}
\begin{aligned}
ds^{2} &=  \frac{1}{z^{2}} \left[ -\left(1-z\right)U\left(z\right) dt^{2} + \frac{dz^{2}}{ \left(1-z\right) U\left(z\right) } + dx^{2} + dy^{2} \right],\\
A_{t} &= \mu\left(1-z\right),
\end{aligned}
\end{equation}
where $U\left(z\right) = 1+z+z^{2} -\mu^2 z^3$, and $ A_{a} $ is the gauge field. The asymptotic boundary is $z = 0$ and the horizon locates at $z=1$. AdS-RN black hole is a
two-parameter system $ \left(\tilde T, \mu\right) $ with $\tilde T = \frac{6-\mu^{2}}{8\pi}$ the Hawking temperature, and $ \mu $ the chemical potential. Moreover, the system is invariant under the rescaling $x_{\alpha} \to \lambda x_{\alpha}$ and $\mu\to\mu/\lambda,\,g_{\#\#} \to g_{\#\#}/\lambda^{2}$. Adopting $ \mu $ as the scaling unit, the scaling-invariant system only has one free parameter $T=\tilde T/\mu$. We shall only focus on scaling-invariant quantities throughout this paper. The scaling-invariant width of a strip and HEE are $w\equiv \tilde w \mu$ and $S \equiv \tilde S /\mu$, respectively. Note that $\tilde w = \int dx$ and $\tilde S$ represent the dimensionfull width and HEE, respectively. In this paper, we label the dimensionful quantities with tilded symbols, while the dimensionless quantities are labeled as symbols without a tilde.

\begin{figure}
\includegraphics[width=0.6\textwidth]{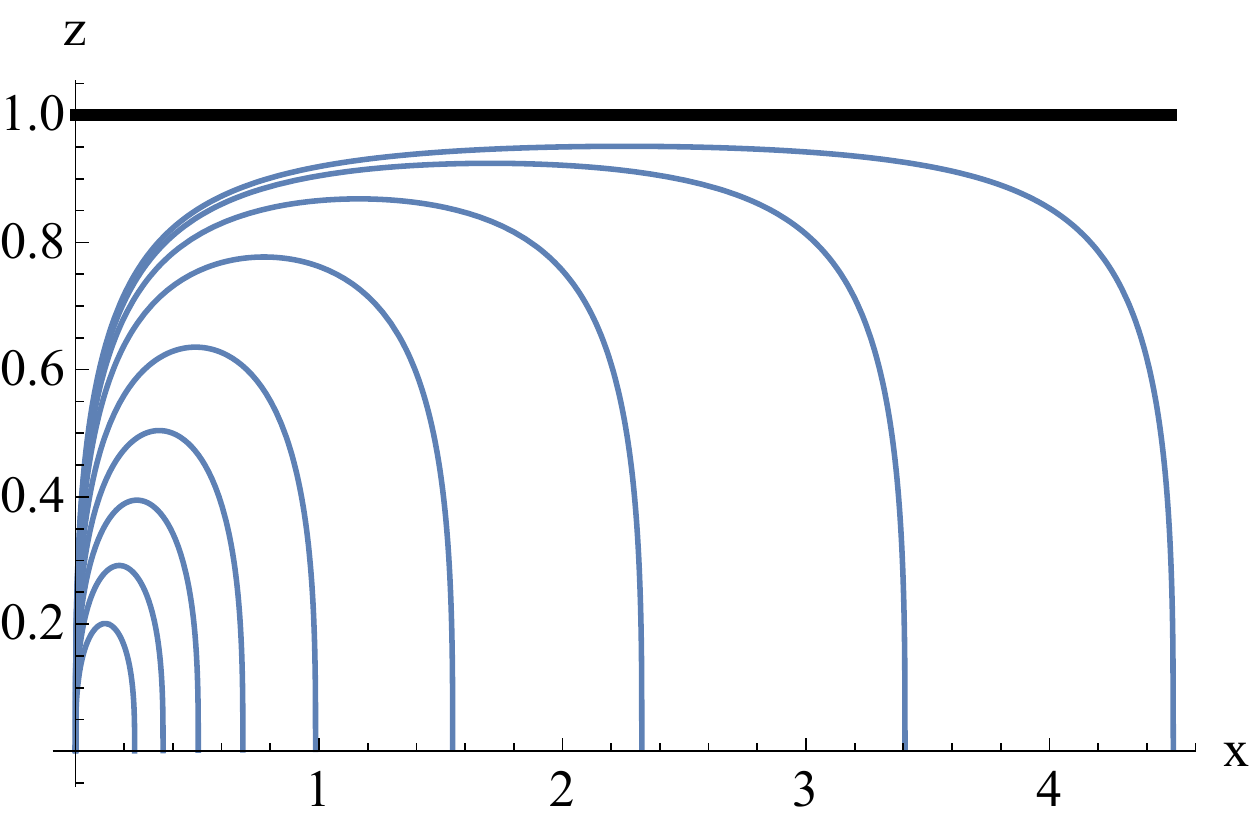}
\caption{Each blue curve is $z(x)$ with different widths, and the black horizontal line is the horizon of AdS-RN black hole. Apparently the curve approaches the horizon as $w$ increases.}\label{fig:zxshow1}
\end{figure}

The minimum surface in AdS-RN black hole has to be solved case by case since the global scaling symmetry is lost. For a generic black hole system, the minimum surface approaches the horizon and becomes more singular as the width of the strip increases (Fig. \ref{fig:zxshow1}), which poses two difficulties for solving the minimum surface. First, the numerical computation of geodesic and other related quantities could fail easily due to the coordinate singularity at the horizon. This difficulty can be overcome by implementing the following coordinate transformation in radial direction.
\begin{equation}\label{coord_change}
z\to 1-\hat z^{2},
\end{equation}
where $z=0 \leftrightarrow \hat z =1, \, z=1 \leftrightarrow \hat z =0$. Second, the singular behavior of the minimum surface prevents us from solving arbitrarily large minimum surface. Despite the absence of large minimum surface, interesting behaviors can still be revealed by relatively small minimum surfaces.

Next, we study the EoP in three different ranges of configurations: small, intermediate and large configurations. We refer the small configurations to the situations when the AdS controls the leading order of the HEE. The HEE behavior gradually deviates from that of the AdS when increasing the size of the configuration. When the HEE behavior begins to deviate significantly from the HEE behavior of the AdS, we refer the configurations at this time as the intermediate configurations. When further increasing the size of the configuration, we can expect the behavior of HEE at certain size of configurations to be similar to the behavior of HEE at infinitely large configurations, which we refer to large configurations.

The terms ``small/intermediate/large configurations'' can also be described in a more precise way. As stated in the previous paragraph, the size of the subregion is determined by whether the corresponding minimum surface is close to the horizon. This can be expressed by comparing the width $w$ with the horizon radius $r_h$. In this paper, the scaling-invariant horizon radius $ r_h/\mu $ can be solved as,
\begin{equation}\label{eq:rhmu}
	\frac{r_{h}}{\mu} = \frac{1}{6} \left(\sqrt{16 \pi ^2 T^2+3}+4 \pi T \right).
\end{equation}
Therefore, for a configuartion $ (a,b,c) $, we can refer the ``small/intermediate/large configuration'' with
\begin{equation}\label{eq:small_config}
	\left\{
	\begin{aligned}
	w_{b} \ll \frac{r_{h}}{\mu}, \,w_{a\cup b\cup c}\; \ll & \; \frac{r_{h}}{\mu} & \text{small}\\
	w_{b}\ll \frac{r_{h}}{\mu}, \,w_{a\cup b\cup c}\; \sim & \; \frac{r_{h}}{\mu} & \text{intermediate}\\
	w_{b} \sim \frac{r_{h}}{\mu},\,w_{a\cup b\cup c}\; \sim & \; \frac{r_{h}}{\mu} & \text{large}
	\end{aligned}
	\right..
\end{equation}

\subsection{Small Configurations}

The EoP for small configuration is dominated by the asymptotic AdS geometry. The sub-leading terms come from the deviation from AdS, which results from the operator deforming the AdS. 
The deformation effect on EoP of small configuration is thus encoded in sub-leading terms, which can be analyzed by asymptotic expansion.

First we discuss the effect of temperature on HEE for small configurations, from which the behavior of EoP can be deduced. The expression \eqref{lag_arc:1} of AdS-RN black hole differs from the case of AdS$ _{4} $ only at $ g_{zz} $,
\begin{equation}\label{eq:gzz-def}
\delta g_{zz} = \frac{1}{z^{2}} \frac{1}{\left(1-z\right)\left(1 + z + z^{2} - \mu^{2} z^{3}\right)} - \frac{1}{z^{2}} = \frac{z+\mu ^2 (1-z) z}{1-z^3 \left(1+\mu ^2 (1-z)\right)}.
\end{equation}
At width $ \tilde w $, the HEE of RN differs from HEE of AdS as,
{\begin{equation}\label{action-adsrn1}
\begin{aligned}
\delta \tilde S = \tilde S^{(\text{AdS-RN})} - \tilde S^{(\text{AdS})} &= \int_{0}^{\tilde w} \frac{z^{2} z'(x)^{2} \delta g_{zz}}{2 z^2 \sqrt{z'(x)^2+1}}dx + \int_{0}^{\tilde w} \text{EOM}_{\text{AdS}} \delta z(x) dx \\
&= \int_{0}^{\tilde w} \frac{z^{2} z'(x)^{2} \delta g_{zz}}{2 z^2 \sqrt{z'(x)^2+1}}dx,
\end{aligned}
\end{equation}}
where $ \delta z(x) $ is the deformation of the minimum surface $ z(x) $ in response to $ \delta g_{zz} $. {Eq. \eqref{action-adsrn1} indicates that only the metric deformation accounts for the temperature behavior of the HEE.}

\begin{figure}[htbp]
	\centering
	\includegraphics[width=0.75\textwidth]{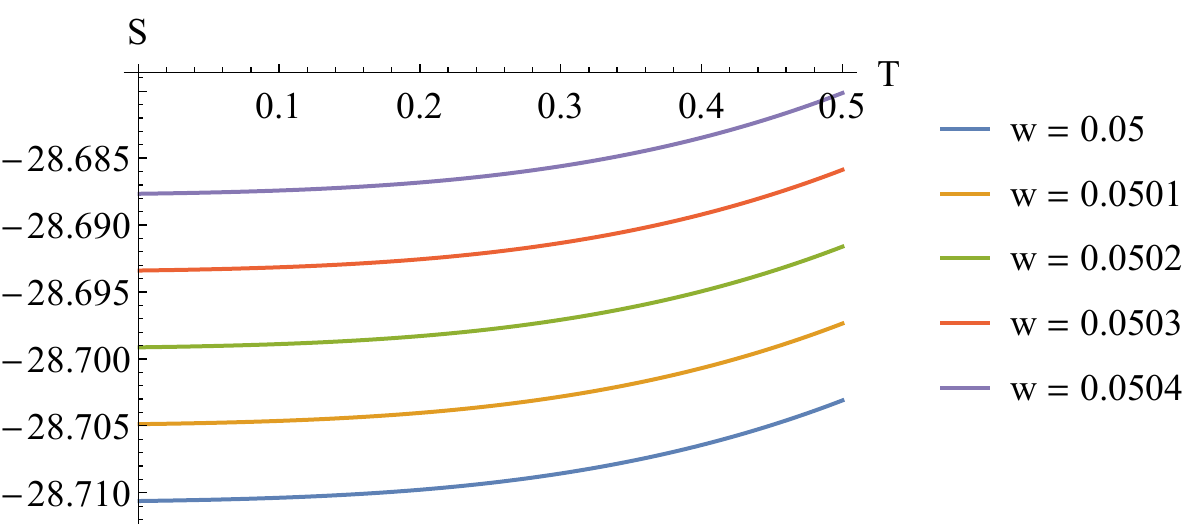}
	\caption{HEE behavior with temperature at different widths.}
	\label{fig:heetrend}
\end{figure}

We now explore the effect of temperature on HEE by studying
\begin{equation}\label{eq:adsrn-deri}
\left.\frac{\partial S}{\partial T}\right|_{w}.
\end{equation}
The HEE for AdS-RN is, 
\begin{equation}\label{eq:adsrn-dif}
S^{(\text{AdS-RN})} = S^{(\text{AdS})} + \delta S.
\end{equation}
It is easily seen that $ \frac{\partial S_{\text{AdS}}}{\partial T} = 0 $, which leaves us with 
\begin{equation}\label{eq:adsrn-dif-expan}
\left.\frac{\partial S}{\partial T}\right|_{w} = \left.\frac{\partial \delta S}{\partial T}\right|_{w} = \pd{\left(\delta \tilde S/\mu\right)}{\mu}{w}\frac{\partial\mu}{\partial T} = \left(\frac{1}{\mu}\pd{\delta \tilde S}{\mu}{w} - \frac{\delta \tilde S}{\mu^{2}}\right) \frac{\partial\mu}{\partial T}.
\end{equation}
Notice also that in AdS-RN black hole,
\begin{equation}\label{eq:dmdt}
\frac{\partial \mu}{\partial T} = 4 \pi \left[\left(1+\frac{3}{8 \pi ^2 T^2}\right)^{-1/2}-1\right] <0.
\end{equation}
$ \delta \tilde S $ is a function of $ \tilde w $ and $ \mu $: $\delta \tilde S = \delta\tilde S(\tilde w,\mu) $, therefore we have
\begin{equation}\label{eq:pdsmu1}
\pd{\delta \tilde S}{\mu}{w} = \pd{\delta \tilde S}{\tilde w}{\mu} \frac{\partial\tilde w}{\partial \mu} + \pd{\delta \tilde S}{\mu}{\tilde w} = - \frac{w}{\mu^{2}} \pd{\delta \tilde S}{\tilde w}{\mu}+ \pd{\delta \tilde S}{\mu}{\tilde w},
\end{equation}
which follows from that $ \pd{\tilde w}{\mu}{w} = \pd{\left(w/\mu\right)}{\mu}{w} = - \frac{ w}{\mu^{2}}$. 
{Therefore, by inserting \eqref{action-adsrn1} we have,}
\begin{equation}\label{eq:dsoverdt}
\begin{aligned}
\left.\frac{\partial S}{\partial T}\right|_{w} &= \left(- \frac{w}{\mu^{2}} \pd{\delta \tilde S}{\tilde w}{\mu}+ \pd{\delta \tilde S}{\mu}{\tilde w}- \frac{\delta \tilde S}{\mu^{2}}  \right)\frac{\partial\mu}{\partial T}\\
& = \left[-\frac{w\left(\mu ^2/2+1\right)}{\mu^{3}} \frac{\partial\Omega}{\partial \tilde w} + \left(\frac{1}{2}-\frac{1}{\mu ^2}\right) \Omega + O\left(z^2\right)\right]\frac{\partial\mu}{\partial T}.
\end{aligned}
\end{equation}
with $\Omega\equiv \int \frac{z(x) z'(x)^2}{2\sqrt{z'(x)^2+1}} dx >0$, and $ O\left(z^{2}\right) $ represents the contribution from the second order expansion of $ z $. The integral $ \Omega \sim [ \tilde w ]^{2} $, therefore $ \frac{\partial\Omega}{\partial \tilde w} = 2 \Omega/\tilde w $. Hence \eqref{eq:dmdt} and \eqref{eq:dsoverdt} leads to that,
\begin{equation}\label{eq:dsoverdtfinal}
\left.\frac{\partial S}{\partial T}\right|_{w} = - \Omega\left(\frac{1}{2} + \frac{3}{\mu^{2}}  \right)  \frac{\partial\mu}{\partial T} > 0.
\end{equation}
Therefore we arrive at the conclusion that $ \partial_{T} S > 0 $. This is also testified by numerical results in Fig. \ref{fig:heetrend}.

Now we point out that EoP for small configurations monotonically increase with temperature as well. The EoP for small configurations of AdS-RN black hole can be expanded as
\begin{equation}\label{ew-cor-1}
E_{W}^{(\text{AdS-RN})} = E_{W}^{\text{(AdS)}} + \delta E_{W},
\end{equation}
where $ \delta E_{W} $ is the correction of AdS-RN to pure AdS$ _{4} $. The geodesic for small configurations in AdS-RN can be seemed as unchanged compared with AdS case, as we argued above. Therefore, the $\delta E_{W} $ can be expanded as
\begin{equation}\label{ew-cor-2}
\delta E_{W} = \delta E_{W}^{(\text{csd})} + \delta E_{W}^{(\text{md})},
\end{equation}
where $ \delta E_{W}^{(\text{csd})} $ is the contribution from deformation of the minimum cross-section, and $ \delta E_{W}^{(\text{md})} $ is the contribution from the metric deformation. Since $ E_{W}^{(\text{AdS})} $ is the area of the minimum cross-section, any deformation to the cross-section will only increase the EoP, therefore $ \delta E_{W}^{(\text{csd})} \geqslant 0 $. To minimize the $ \delta S $ is to take $ \delta E_{W}^{(\text{csd})} = 0$, \emph{i.e.}, the minimum cross-section of AdS-RN is the same as that of AdS$_{4}$. Therefore the $ \delta E_{W} $ comes only from the metric deformation. {That is to say, we only need to study the influence of the metric deformation on the area of minimum cross-section (one line segment) in the AdS space. Notice that the HEE is the area of the minimum surface, and we also proved that the HEE monotonically increases with temperature under the influence of the metric deformation. Therefore, following this argument, we arrive at the conclusion that EoP for small configurations monotonically increase with temperature.}

{Throughout this paper we focus on the scale-invariant quantities. 
Some recent studies on dimensional EoP have come to different conclusions from our paper. This is actually as expected, as the scale invariance is essential to our conclusions.
For example, if we focus only on dimensionful quantities, Eq. \eqref{eq:dsoverdtfinal} becomes,
\begin{equation}\label{eq:dsoverdtfinal_2}
\left.\frac{\partial S}{\partial T}\right|_{w} = \frac{\Omega}{2}\frac{\partial\mu}{\partial T} < 0.
\end{equation}
Therefore, the dimensionful EoP now {\it decrease} with temperature, following the arguments of deducing the temperature behavior of dimensionless EoP.}

The above analysis can be directly applied to other black hole systems, because the deformation of the AdS black hole can be studied by the asymptotic expansion.
Despite the simple monotonic behavior in AdS-RN black hole, the EoP behavior with system parameter for small configurations could be more diverse in other holographic models.

Asymptotic expansion does not apply to the intermediate configurations, so we can only study it directly using numerical methods.

\subsection{Intermediate Configurations}
Next we numerically compute the intermediate configurations EoP for AdS-RN black hole. We intend to investigate the behavior of EoP with temperature at a fixed configuration $(a,b,c)$ and then consider the dependence of EoP on configuration parameters at fixed temperature.

\begin{figure}[htbp]
	\centering
	\includegraphics[width=0.45\textwidth]{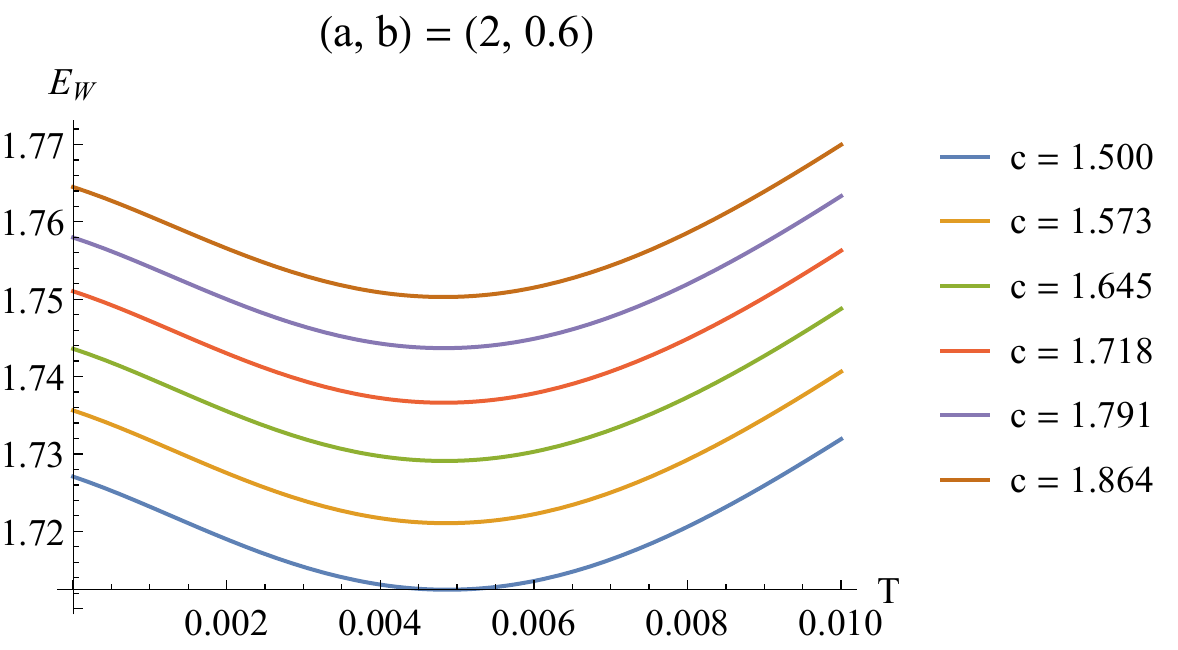}\quad
	\includegraphics[width=0.45\textwidth]{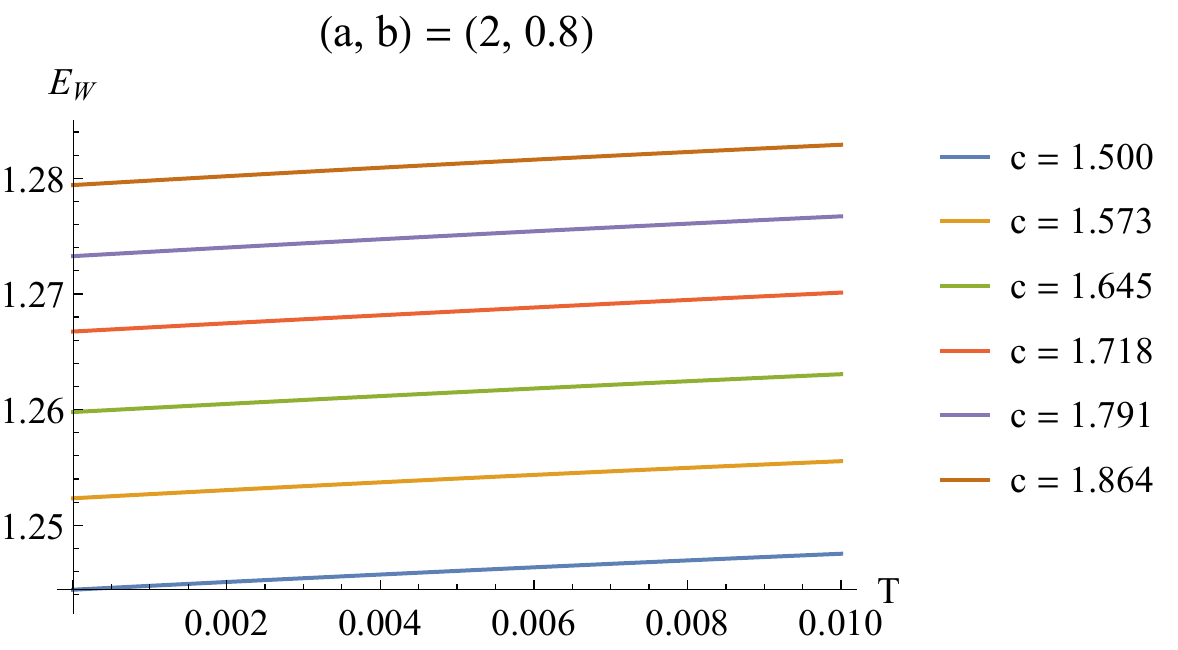}
	\caption{EoP $ E_{W} $ vs $ T $ at various $ (a,b,c) $. Different curves correspond to different $ c $ at $ (a,b)=(2,0.6)$ (left plot) and $ (a,b) = (2,0.8) $ (right plot).}
	\label{fig:diffTandabc}
\end{figure}

Despite the monotonic temperature behavior for small configurations, 
the intermediate configuration EoP presents more diverse phenomena.
Fig. \ref{fig:diffTandabc} shows $ E_{W} $ vs $ T $ at different configurations. For $ (a,b)=(2,0.6) $, the EoP first decrease with temperature and then increase with temperature (see the left plot of Fig. \ref{fig:diffTandabc}); while for $ (a,b) = (2,0.8) $ EoP increases with temperature monotonically (see the right plot of Fig. \ref{fig:diffTandabc}). Therefore the temperature behavior is configuration-dependent. The reason for configuration dependent EoP behavior is that the definition of EoP itself is complicated. Like many other quantum information-related quantities, there may be a complex relationship between EoP itself and system parameters \cite{Amico:2007ag}.

Another interesting phenomenon is that the temperature behavior of EoP is more sensitive to the value of $ b $ than the value of $ a $ and $ c $. 
{We can see from Fig. \ref{fig:diffTandabc} that the temperature behavior of the different curves (corresponding to different $c$ values) in each plot are similar. However, by directly comparing the left and right plot (where $b$ values are different), it can be found that their temperature behavior are different.} This is expected since the EoP is mainly contributed from the region near the bottom of $ C_{b} $. During phase transitions, however, the EoP could be sensitive to $ a $ and $ c $ since phase transition are usually accompanied by deformations of near horizon geometry, at which the $ C_{a,b,c} $ locates.

\begin{figure}[htbp]
	\centering
	\includegraphics[width=0.45\textwidth]{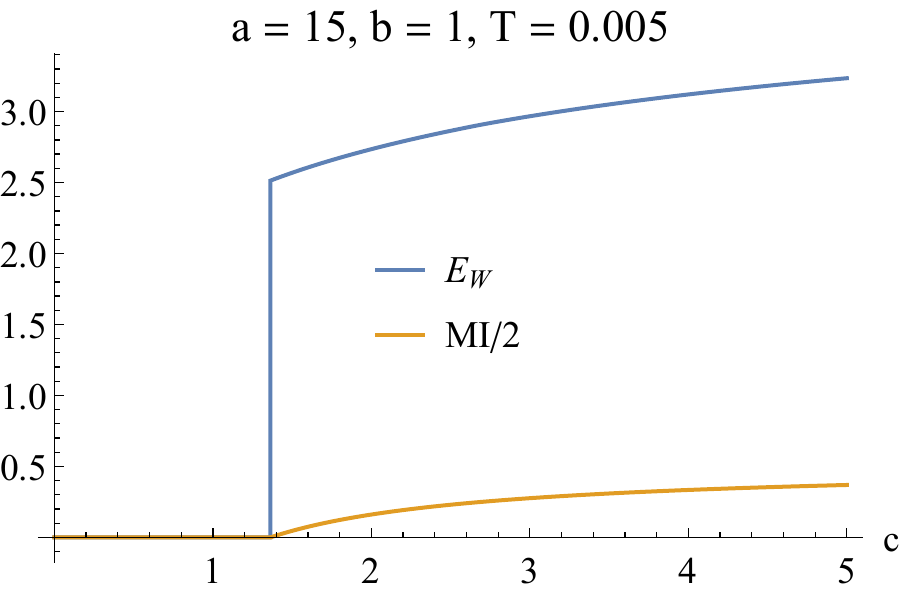}\quad
	\includegraphics[width=0.45\textwidth]{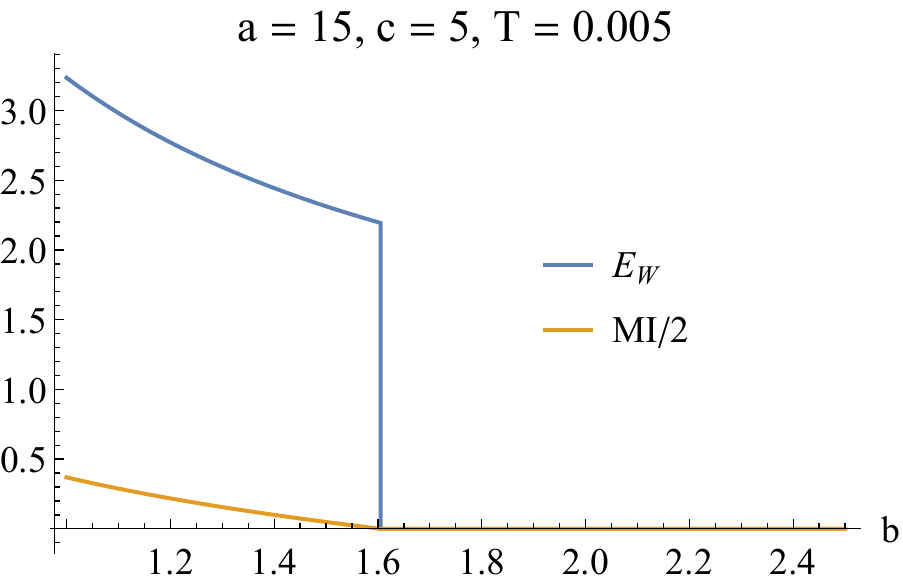}
	\caption{EoP and MI$ /2 $ at different configurations. Left plot: $ E_{W} $ vs $ c $ at $ (a,b,T) = (15,1,0.005) $. Right plot: $ E_{W} $ vs $ b $ at $ (a,c,T) = (15,5,0.005) $.}
	\label{fig:difctempic}
\end{figure}

Next we study the EoP dependence on configurations (see Fig. \ref{fig:difctempic}). Again, we find that $ E_{W} $ increases (decreases) with $ c $ ($ b $), and EoP is always greater than one half of the MI. The disentangling phase transition of EoP can also be observed when MI starts to vanish.

\subsection{Large Configurations}

For large configurations, where $ (a,b,c) $ are all large,  the EoP vanishes as the MI vanishes. The geodesics for large subregions are close to the horizon (see Fig. \ref{fig:zxshow1}), hence the HEE will be dominated by the thermal entropy. Subsequently, the MI for large configurations must vanish\footnote{{For large configuration limit, all of $ C_{a}, C_{b},C_{c},C_{a,b,c} $ are close to the horizon and the HEE are dictated by the thermal entropy. Consequently, $ S(A\cup B\cup C) + S(B) \simeq S(A) + 2S(B) + S(C) > S(A) + S(C) $. Therefore we have $ S(A\cup C) = S(A)+S(C) $ and $ I(A,C) = 0 $. Strictly speaking, we should also take into account the area of the geodesic from boundary to the horizon, since the HEE for large configurations are contributed from the near horizon region and the straight line from the boundary to the horizon. But this contribution is small compared to the thermal entropy for large configurations, and hence can be neglected.}} and result in vanishing EoP. This property also naturally results from statistical mechanics. 
For large sub-regions, EE is mainly contributed by thermal entropy because thermal entropy exhibits volume law, while EE exhibits area law.
Therefore the density matrix of two separate large subsystems tends to be a product of thermal states of each subsystem, and leads to the vanishing of MI and EoP.

The EoP and MI also vanishes in high temperature limit. The high temperature limit $\mu\to 0$ indicates that a finite $ w $ corresponds to an infinite $ \tilde w $. Therefore $C_{a},C_{b},C_{c},C_{a,b,c}$ for finite $\left(a,b,c\right)$ are all close to the horizon, and the corresponding HEE are again dominated by thermal entropy. The MI and EoP will vanish, following the explanation of the large configuration limit. Therefore it is the same gravitational nature that leads to the vanishing EoP in high temperature limit, and in large configuration limit. From the viewpoint of statistical mechanics, the total density matrix of bipartite systems at large temperatures can be approximated as direct products of the thermal density matrix of each subsystem. Therefore, the EoP and MI will vanish in high temperature limit.

\section{Discussion}\label{discussion}

We have investigated the EoP for general strip configurations in AdS$_{4}$ spacetime and AdS-RN black hole in this paper. In both cases we have found that EoP increases (decreases) with subregions (separation), and EoP is greater than half of the MI. For AdS$_{4}$ the scaling symmetry simplifies the computation. For AdS-RN black hole we study the EoP behavior for three different ranges of configurations: the EoP monotonically increase with temperature for small configurations; for intermediate configurations the temperature behavior of EoP depends on configurations; for large configuration, and also for high temperature limit, the EoP vanishes. Our work offers a general discussion on EoP in holographic black hole systems, which can inspire more investigations in the future. Next, we point out several topics worthy of further exploration.

Using the techniques developed in this paper, we can study the EoP in more general holographic systems. First, the discussion on EoP for AdS$_{4}$ can be immediately generalized to general AdS$_{d}$. Second, our algorithm can be applied to a general multi-partition configuration on any homogeneous background. Moreover, general configuration EoP is also worthy to study, but this usually involves in solving complicated partial differential equations. More importantly, the intimate connection between entanglement and physics suggests that EoP is closely related to the physical properties of holographic systems. For example, the HEE exhibits interesting phenomena during a
thermal phase transition \cite{Cai:2015cya}. It can be expected that EoP in these thermal phase transitions will also have important applications. For quantum phase transitions, we can expect that the system will exhibit novel behaviors in zero temperature limit, such as the scaling behavior of EoP in critical region. EoP can also be used to explore the properties of dynamic systems. We plan to explore above directions in the future.

\section*{Acknowledgments}

We are very grateful to Long Chen, Wu-Zhong Guo, Peng-Xu Jiang, Wei-Jia Li for helpful discussions and suggestions. Peng Liu would like to thank Yun-Ha Zha for her kind encouragement during this work. This work is supported by the Natural Science Foundation of China under Grant No. 11575195, 11875053, 11805083, 11847055, 11905083 and 11775036.

\begin{appendix}
\section{Geometrical proof of inequalities related to EoP}\label{sec:geoprof}

In \cite{Takayanagi:2017knl} several inequalities related to EoP have been discussed and proved in global coordinates. The satisfaction of these inequalities for EoP is one of the major motivations for the proposal of holographic EoP. Here we prove three important inequalities of EoP directly in Poincar\'e coordinates. These inequalities have been verified by the numerical results for AdS$_{4}$ and AdS-RN black hole as presented in previous sections.

\begin{figure}[htbp]
    \centering
    \includegraphics[width=0.45\textwidth]{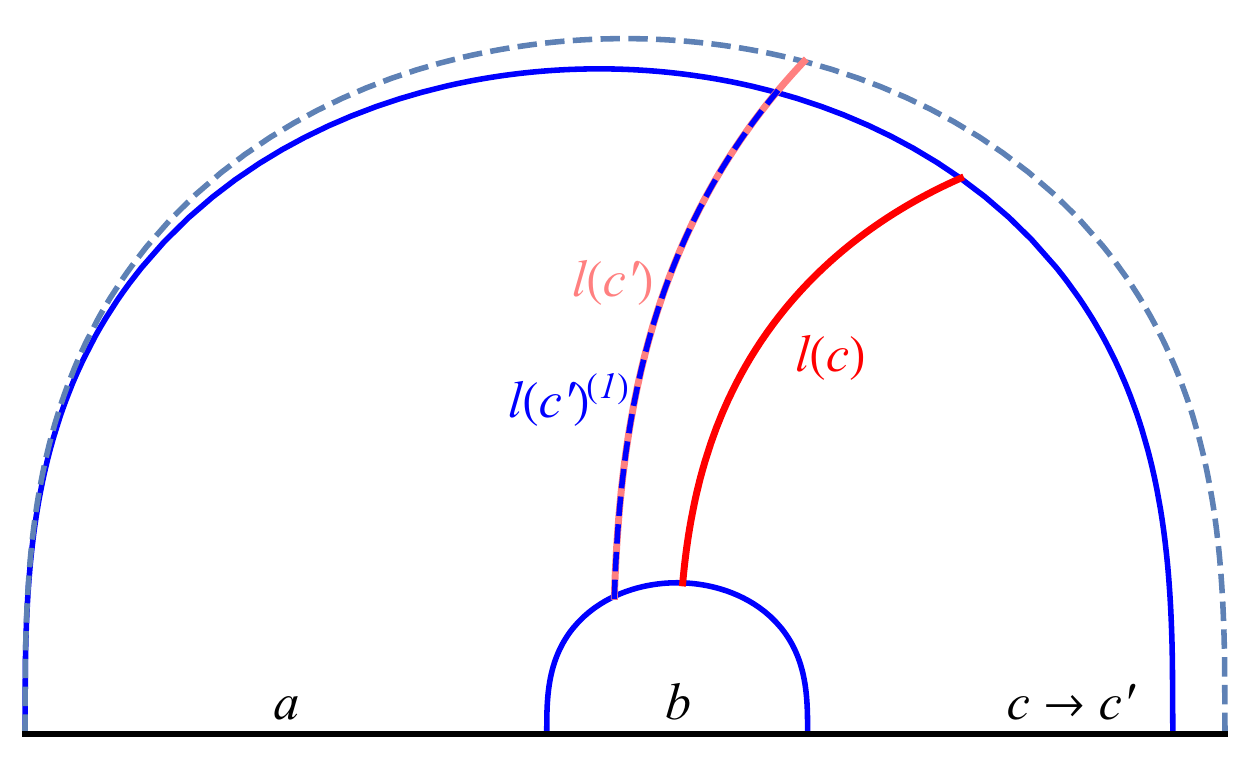}\qquad
    \includegraphics[width=0.45\textwidth]{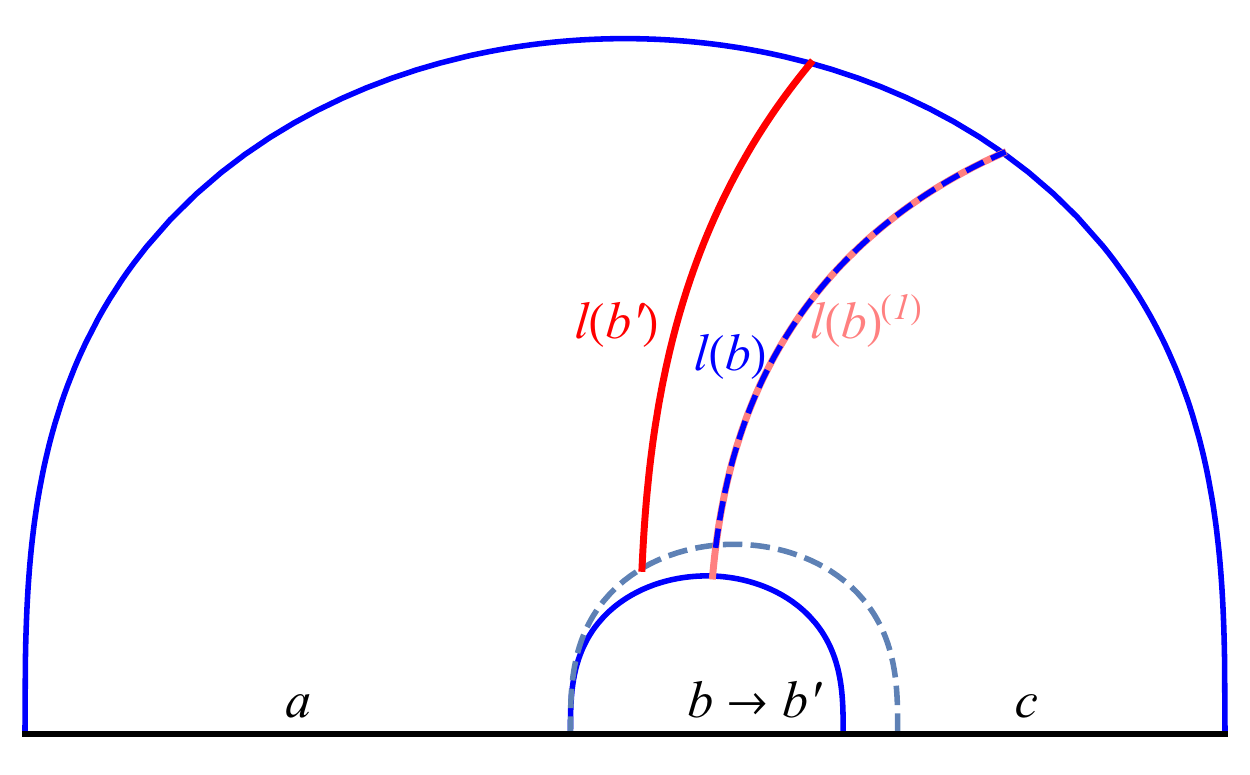}
\caption{Left plot: the cartoon for the proof of inequality \eqref{eq:prove1}. Two blue curves are $C_{b},\,C_{a,b,c}$ respectively, and the dashed curve in light blue is $C_{a,b,c'}$. The red and pink curve are the minimum cross-sections for $\left(a,b,c\right)$, $\left(a,b,c'\right)$ with EoP as $l\left(c\right),\,l\left(c'\right)$, respectively. The blue dashed curve is the segment of the minimum cross-section for $\left(a,b,c\right)$ with length $l\left(c'\right)^{\left(1\right)}$. Right plot: the cartoon for the proof of inequality \eqref{eq:prove2}. Two blue curves correspond to $C_{b},\,C_{a,b,c}$ respectively. The dashed light blue curve is $C_{a,b',c}$. The red and pink curve are the minimum cross-sections for $\left(a,b',c'\right)$, $\left(a,b,c\right)$ with EoP as $l\left(b'\right),\,l\left(b\right)$, respectively. The blue dashed curve is the segment of the minimum cross-section for $\left(a,b,c\right)$ with length $l\left(b'\right)^{\left(1\right)}$.}
    \label{fig:prove12}
\end{figure}

\begin{figure}[htbp]
    \centering
    \includegraphics[width=0.65\textwidth]{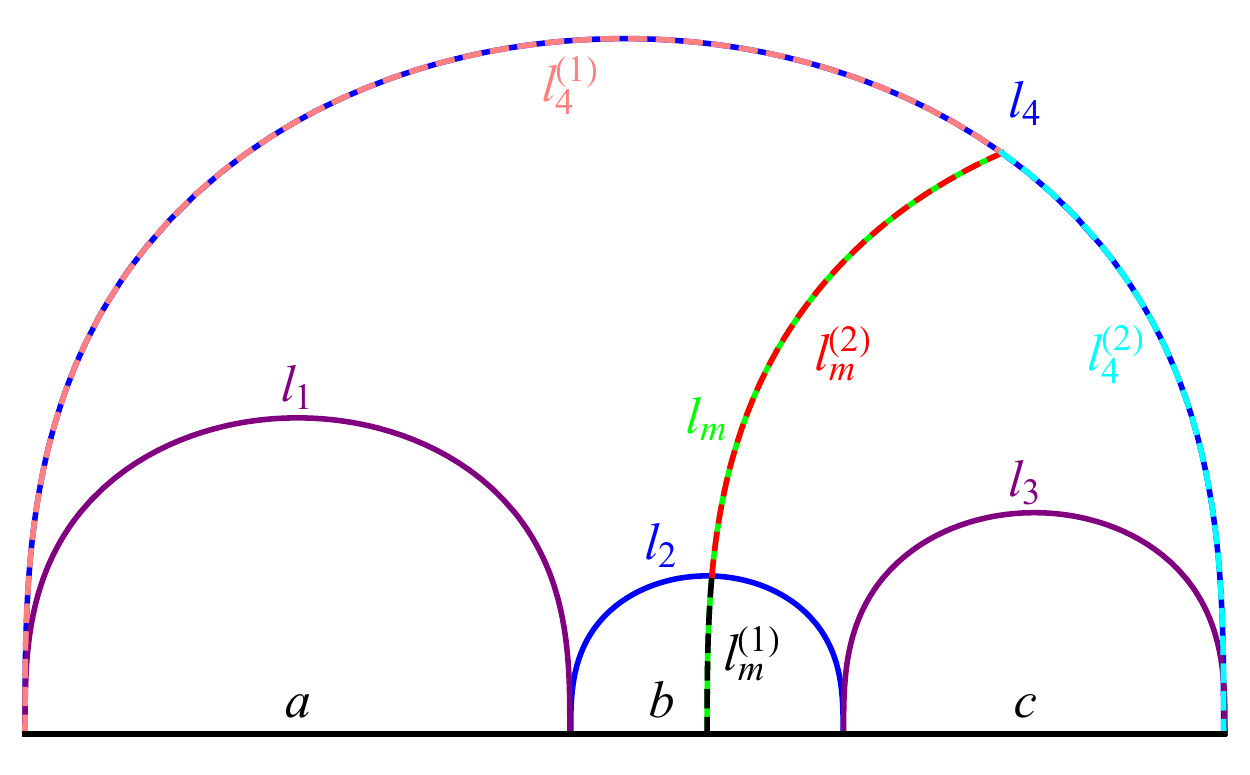}
    \caption{
	A cartoon for the proof of inequality \eqref{eq:mieopineq}.	Two purple curves are minimum surfaces with disconnected entanglement wedge, and two blue curves are minimum surfaces with connected entanglement wedge. $l_{1},l_{2},l_{3},l_{4}$ are the lengths of the curves of the geodesic $C_{a},C_{b},C_{c},C_{a,b,c}$ respectively. The green curve is the geodesic intersecting with $C_{b}$ and $C_{a,b,c}$ from which the minimum cross-section is obtained. The length of the green curve is $l_{m}$ and the EoP of configuration $(a,b,c)$ is $l_{m}^{(2)}$.
    }
    \label{fig:provemi}
\end{figure}

\begin{enumerate}
    \item
    \begin{equation}\label{eq:prove1}
    E_{W}\left(a,b,c+\delta c\right)\geqslant
    E_{W}\left(a,b,c\right) \quad \text{with} \quad\delta c\geqslant 0
    \end{equation}
For a fixed $\left(a,b\right)$, when increasing $c\to c+\delta c$, the $C_{a,b,c+\delta c}$ encapsulates larger region than $C_{a,b,c}$. Therefore, EoP will increase with $c$. The proof is quite transparent, as illustrated in Fig. \ref{fig:prove12}. Suppose $l(c)>l(c')$, then $l(c)>l(c')^{(1)}$, which contradicts with the fact that $l(c)$ is the minimum cross-section. Therefore, there must be $l(c)\leqslant l(c')$. This is equivalent to $E_{W}\left(\rho_{A\left(BC\right)}\right)\geqslant E_{W}\left(\rho_{AB}\right)$ as discussed in \cite{Takayanagi:2017knl}.
    \item
    \begin{equation}\label{eq:prove2}
    E_{W}\left(a,b+\delta b,c-\delta b\right)\leqslant
    E_{W}\left(a,b,c\right) \quad \text{with} \quad\delta b\geqslant 0
    \end{equation}
This inequality says that when increasing $b$ to $b'$ with fixed $a+b+c$, the entanglement wedge of $(a,b',c)$ is smaller than that for $(a,b,c)$. Therefore, EoP will decrease with increasing $b$. If $l(b')>l(b)$, then $l(b')>l(b)^{(1)}$, which contradicts with the fact that $l(b')$ is the minimum cross-section.
    \item The relation to MI:
    \begin{equation}\label{eq:mieopineq}
    E_{W}\left(\rho_{AB}\right)=l^{(2)}_{m}\geqslant \frac{I\left(A:B\right)}{2} = \frac{1}{2}\left(l_{1} + l_{3} -l_{2} - l_{4}\right).
    \end{equation}
    The EoP $E_{W}(a,b,c)=l^{(2)}_{m}$ (the length of the red dashed curve in Fig. \ref{fig:provemi}), which is a segment of $l_{m}$. Then we have the following relation,
    \begin{equation}\label{eqnpro1}
    E_{W}\left(\rho_{AB}\right)=l_{m}^{(2)}=l_{m} - l_{m}^{(1)} \geqslant l_m - \frac{l_{2}}{2}= \left( l_{m} + \frac{l_{4}}{2} \right) - \frac{l_{4}}{2} - \frac{l_{2}}{2}.
    \end{equation}
    The first inequality in \eqref{eqnpro1} is derived from $l_{m}^{1}\leqslant l_{2}/2$, which can be easily proved. Therefore the proof is completed if $l_{m} + \frac{l_{4}}{2} \geqslant \frac{l_{1} + l_{3}}{2}$. This is readily seen if we break the $l_{4}$ into $l^{\left(1\right)}_{4}$ and $l^{\left(2\right)}_{4}$. It is seen that $l^{\left(1\right)}_{4} + l_{m} \geqslant l_{1},\; l^{\left(2\right)}_{4} + l_{m} \geqslant l_{3}$. Therefore \eqref{eq:mieopineq} is proved.
\end{enumerate}

\end{appendix}

\end{document}